\documentclass[10pt]{amsart}

\usepackage{lipsum}
\usepackage{amsfonts,bm,amssymb,amsmath}
\usepackage{graphicx}
\usepackage{epstopdf}
\usepackage[caption=false]{subfig}
\usepackage{pgfplots}
\usepackage{algorithmic}
\usepackage{amsopn}
\usepackage{amsrefs}
\usepackage{float}
\usepackage{caption}

\usepackage{hyperref}
\usepackage{cleveref}

\usepackage{color}
\usepackage{diagbox}

\usepackage{geometry}

\newtheorem{remark}{Remark}[section]

\hypersetup{
	colorlinks,
	linkcolor={red!50!black},
	citecolor={blue!50!black},
	urlcolor={blue!80!black}
}

\numberwithin{equation}{section}

\definecolor{newcolor1}{rgb}{.8,.349,.1}
\colorlet{bblue}{blue!50!black}
\crefformat{equation}{(#2#1#3)}
\crefmultiformat{equation}{(#2#1#3)}{ and~(#2#1#3)}{, (#2#1#3)}{ and~(#2#1#3)}

\crefformat{figure}{Figure~#2#1#3}
\crefmultiformat{figure}{Figures~ #2#1#3}{ and~#2#1#3}{, (#2#1#3)}{ and~(#2#1#3)}
\crefformat{table}{Table~#2#1#3}

\def\a{\mbox{\boldmath $a$}}

\def\e{\mbox{\boldmath $e$}}
\def\f{\mbox{\boldmath $f$}}

\def\g{\mbox{\boldmath $g$}}
\def\h{\mbox{\boldmath $h$}}
\def\m{\mbox{\boldmath $m$}}

\def\x{\mbox{\boldmath $x$}}
\def\y{\mbox{\boldmath $y$}}

\def\0{\mbox{\boldmath $0$}}

\begin{document}

\title[A novel third-order method for micromagnetics]{A novel third-order accurate and stable scheme for micromagnetic simulations}

\author[C. Xie]{Changjian Xie}
\address{School of Mathematics and Physics\\ Xi'an-Jiaotong-Liverpool University\\Re'ai Rd. 111, Suzhou, 215123, Jiangsu\\ China.}
\email{Changjian.Xie@xjtlu.edu.cn}



\date{\today}


\begin{abstract}
High-fidelity numerical simulation serves as a cornerstone for exploring magnetization dynamics in micromagnetics. This work introduces a novel third-order temporally accurate and stable numerical scheme for the Landau-Lifshitz-Gilbert (LLG) equation, aiming to address the limitations in accuracy and efficiency often encountered with conventional approaches. Validation via nanostrip simulations confirms two principal advantages of the proposed method: it attains strict third-order temporal accuracy, surpassing many current techniques, and it offers superior computational efficiency, enabling rapid convergence without sacrificing numerical precision. For Gilbert damping coefficients $\alpha$ ranging from $0.1$ to values below $10$, the scheme preserves strong stability and effectively avoids non-physical magnetization states. The magnetic microstructures predicted by this method are in excellent agreement with those from established benchmark methods, affirming its reliability for quantitative physical analysis. Salient distinctions between the proposed scheme and an existing third-order semi-implicit method include:
\begin{itemize}
	\item[(1) ] Solving the linear system associated with the existing scheme demands substantially greater computational time, underscoring the need for highly efficient solvers;
	\item[(2) ] Although the proposed method shows increased sensitivity to damping parameters, it reliably converges to stable physical states and is effective in simulating magnetic domain wall motion, producing outcomes consistent with prior validated studies;
	\item[(3) ] The energy levels computed by the proposed method are significantly lower than those obtained via the existing third-order scheme.
\end{itemize}
\end{abstract}

\maketitle

\section{Introduction}

Ferromagnetic materials find widespread application in data storage technologies, largely due to the bistable nature of their intrinsic magnetic ordering, commonly referred to as magnetization. The temporal evolution of magnetization in these systems is principally dictated by the Landau-Lifshitz-Gilbert (LLG) equation \cite{Landau1935On,Gilbert:1955}. This equation incorporates two fundamental terms governing magnetization dynamics: a gyromagnetic term, responsible for energy conservation, and a damping term, which models energy dissipation. The damping term is critically important in magnetic devices, profoundly influencing both their energy consumption and operational speed. Recent experimental work on magnetic-semiconductor heterostructures \cite{Zhang2020ExtremelyLM} has confirmed that the Gilbert damping constant can be tuned. At the microscopic scale, damping arises from various physical processes, including electron scattering, itinerant electron relaxation \cite{Heinrich1967TheIO}, and phonon-magnon coupling \cite{Suhl1998TheoryOT, Nan2020ElectricfieldCO}, which can be quantitatively assessed through electronic structure calculations \cite{TangXia2017}. From an applied standpoint, modulating the damping parameter enables optimization of magnetodynamic properties in ferromagnetic materials—for instance, by lowering the switching current and enhancing the writing speed of magnetic memory devices \cite{Wei2012MicromagneticsAR}. Although many experimental studies have focused on systems with small damping parameters \cite{Budhathoki2020LowGD,Lattery2018LowGD,Weber2019GilbertDO}, substantial damping effects have also been reported in works like \cite{GilbertKelly1955, Tanaka2014MicrowaveAssistedMR}. Notably, Tanaka et al. \cite{Tanaka2014MicrowaveAssistedMR} found that larger damping constants correlate with reduced magnetization switching times. Gilbert and Kelly \cite{GilbertKelly1955} also reported observing very large damping parameters (around 9) in their investigations. The LLG equation constitutes a vector-valued, nonlinear system characterized by a point-wise constant magnitude constraint on the magnetization vector. Significant research efforts have been dedicated to devising efficient and numerically stable methods for micromagnetic simulations, as summarized in review articles such as \cite{kruzik2006recent,cimrak2007survey}. Semi-implicit schemes have attracted considerable interest among existing numerical strategies due to their ability to maintain good numerical stability without requiring complex nonlinear solvers \cite{alouges2006convergence, gao2014optimal, Xie2018}. For instance, our research group previously proposed a second-order backward differentiation formula (BDF2) scheme based on one-sided interpolation \cite{Xie2018}. This method requires solving a three-dimensional linear system with non-constant coefficients at each time step. A rigorous theoretical analysis establishing the second-order convergence of this BDF2 method was provided in \cite{jingrun2019analysis}. Alternatively, Alouges et al. \cite{alouges2006convergence}  introduced a linearly implicit method utilizing the tangent space to preserve the magnetization constraint, although this approach achieves only first-order temporal accuracy. More recently, Lubich et al. \cite{Lubich2021} developed and analyzed high-order BDF schemes for the LLG equation. Unconditional unique solvability for such semi-implicit schemes has been established in \cite{jingrun2019analysis,Lubich2021}; however, their convergence analysis typically requires the temporal step size to be proportional to the spatial grid size. Despite these developments, the practical implementation and evaluation of high-order numerical methods for micromagnetic modeling continue to be active research areas. A notable limitation persists: most existing methods are limited to first or second-order temporal accuracy. Furthermore, there is a scarcity of third-order accurate methods tailored for practical micromagnetic simulations, especially those capable of handling arbitrary damping parameters and facilitating quantitative comparisons among different numerical approaches. To address this gap, this paper presents a third-order accurate numerical scheme for solving the LLG equation with arbitrary damping parameters. We also conduct comprehensive computational tests to verify the numerical stability of the proposed method. These tests further reveal that the dissipation properties of the new scheme differ from those of our earlier method \cite{xie2025schemeB} when applied to systems with large damping parameters.

The rest of this paper is organized as follows. \cref{sec: numerical scheme} begins with a review of the micromagnetic model, followed by a detailed description of the proposed numerical scheme and a comparative discussion with first-order (BDF1) and second-order (BDF2) semi-implicit projection methods, as well as an existing third-order semi-implicit scheme. \cref{sec:experiments} presents extensive numerical results, encompassing verification of temporal and spatial accuracy in one-dimensional (1D) and three-dimensional (3D) settings, assessments of computational efficiency (via comparisons with BDF1 and BDF2), stability analysis with respect to the damping parameter, and an investigation of domain wall velocity dependence on both damping and external magnetic field. Concluding remarks and potential future research directions are provided in \cref{sec:conclusions}.

\section{The physical model and the numerical scheme}
\label{sec: numerical scheme}

\subsection{Governing equation}

The Landau-Lifshitz-Gilbert (LLG) equation forms the fundamental basis of micromagnetics, providing a rigorous description of the spatiotemporal evolution of magnetization in ferromagnetic materials by incorporating two key physical phenomena: gyromagnetic precession and dissipative relaxation \cite{Landau1935On,Brown1963micromagnetics}. In nondimensional form, this governing equation is expressed as
\begin{align}\label{c1-large}
{\m}_t =-{\m}\times{\bm h}_{\text{eff}}-\alpha{\m}\times({\m}\times{\bm h}_{\text{eff}})
\end{align}
subject to the homogeneous Neumann boundary condition
\begin{equation}\label{boundary-large}
\frac{\partial{\m}}{\partial {\bm \nu}}\Big|_{\partial \Omega}=0,
\end{equation}
where \(\Omega \subset \mathbb{R}^d\) (\(d=1,2,3\)) represents the bounded domain of the ferromagnetic material, and \(\bm \nu\) is the unit outward normal vector on the boundary \(\partial \Omega\). This boundary condition ensures no magnetic surface charge, a physically appropriate assumption for isolated ferromagnetic systems.

The magnetization field \(\m: \Omega \to \mathbb{R}^3\) is a three-dimensional vector field satisfying the pointwise constraint \(|\m| = 1\),  stemming from the quantum mechanical alignment of electron spins in ferromagnets. The first term on the right-hand side of \cref{c1-large} describes gyromagnetic precession, where magnetic moments precess around the effective field \(\bm h_{\text{eff}}\). The second term represents dissipative relaxation, with \(\alpha > 0\) being the dimensionless Gilbert damping coefficient that governs the rate of energy dissipation into the lattice.

From the perspective of the Gibbs free energy functional, the effective field \(\bm h_{\text{eff}}\) is obtained as the functional derivative of the Gibbs free energy \(F[\m]\) with respect to the magnetization, i.e., \(\bm h_{\text{eff}} = -\delta F/\delta \m\). his functional incorporates all relevant energy contributions in ferromagnetic systems—exchange, anisotropy, magnetostatic (stray), and Zeeman energies—and is given by
\begin{equation}\label{LL-Energy}
F[\m] = \frac {\mu_0 M_s^2}{2} \left\{\int_\Omega \left( \epsilon|\nabla\m|^2 +
q\left(m_2^2 + m_3^2\right)
-2\h_e\cdot\m - \h_s\cdot\m \right)\mathrm{d}\x \right\} . 
\end{equation}
Here, \(\mu_0 = 4\pi \times 10^{-7}\, \text{H/m}\) is the vacuum permeability, \(M_s\) is the saturation magnetization, and \(\epsilon\) and \(q\) are dimensionless parameters defined subsequently. The vectors \(\h_e\) and \(\h_s\) denote the externally applied magnetic field and the stray field, respectively. For uniaxial ferromagnetic materials with a single easy axis, the effective field \(\bm h_{\text{eff}}\) decomposes into distinct physical components, yielding the explicit expression
\begin{align}
{\bm h}_{\text{eff}} =\epsilon\Delta\m-q(m_2\e_2+m_3\e_3)+\h_s+\h_e,
\end{align}
where \(\epsilon = C_{\text{ex}}/(\mu_0 M_s^2 L^2)\) and \(q = K_u/(\mu_0 M_s^2)\). Here, \(L\) is the characteristic length scale, \(C_{\text{ex}}\) is the exchange constant (controlling short-range spin alignment), and \(K_u\) is the uniaxial anisotropy constant (representing the energy cost for deviation from the easy axis). The unit vectors \(\e_2 = (0,1,0)\) and \(\e_3 = (0,0,1)\) efine the hard axes perpendicular to the uniaxial easy axis, and \(\Delta\) is the Laplacian operator in \(d\)-dimensional space.
For Permalloy (NiFe), a common soft ferromagnetic material in spintronics, standard parameter values from the literature are: \(C_{\text{ex}} = 1.3 \times 10^{-11}\, \text{J/m}\), \(K_u = 100\, \text{J/m}^3\), and \(M_s = 8.0 \times 10^5\, \text{A/m}\). The stray field \(\h_s\) originates from magnetic charge distributions at domain boundaries and material surfaces and is mathematically represented by the integral expression
\begin{align}\label{eqn:div}
{\h}_{\text{s}}=\frac{1}{4\pi}\nabla \int_{\Omega} \nabla\left( \frac{1}{|\x-\y|}\right)\cdot {\bm m}({\bm y})\,d{\bm y},
\end{align}
which is a formulation that exhibits long-range spatial correlations. A critical computational advancement for practical micromagnetic simulations is that for rectangular domains \(\Omega\), the evaluation of \(\h_s\) can be efficiently computed via the Fast Fourier Transform (FFT) \cite{Wang2000}, which reduces the asymptotic computational complexity from \(O(N^d)\) to \(O(N^d \log N)\) for \(d\)-dimensional grids, enabling large-scale simulations.

To facilitate numerical discretization, we introduce the composite source term
\begin{align}\label{eq-4}
\f=-Q(m_2\e_2+m_3\e_3)+\h_s+\h_e.
\end{align}
which aggregates the anisotropy, stray field, and external field contributions. Substituting this source term into \cref{c1-large}, the LLG equation is re-expressed as
\begin{align}\label{eq-5}
\m_t=-\m\times(\epsilon\Delta\m+\f)-\alpha\m\times\m\times(\epsilon\Delta\m+\f).
\end{align}
Leveraging the vector triple product identity \(\a \times ({\bm b} \times {\bm c}) = (\a \cdot {\bm c}){\bm b} - (\a \cdot {\bm b}){\bm c}\) and the pointwise constraint \(|\m| = 1\) (which implies \(\m \cdot \partial_t \m = 0\) via time differentiation), we simplify \cref{eq-5} to an equivalent formulation that is more amenable to stable numerical discretization:
\begin{equation}\label{eq-model}
\m_t=\alpha  (\epsilon\Delta\m+\f)+\alpha \left(\epsilon |\nabla \m|^2 -\m \cdot\f \right)\m-\m\times(\epsilon\Delta\m+\f).
\end{equation}
We establish a standardized discretization framework to underpin subsequent numerical approximations, defining temporal and spatial discretization notations and boundary condition enforcement strategies.

Let \(k = t^{n+1} - t^n\) denote the uniform temporal step-size, with discrete time levels given by \(t^n = nk\) for \(n = 0,1,\dots,N_T\), where \(N_T = \lfloor T/k \rfloor\) and \(T\) is the final simulation time. For spatial discretization, a uniform Cartesian grid is employed with mesh-size \(h_x = h_y = h_z = h = L/N\) (for cubic domains of characteristic length \(L\)). The notation \(\m_{i,j,\ell}^n\) denotes the numerical approximation of \(\m(x_{i-1/2}, y_{j-1/2}, z_{\ell-1/2}, t^n)\), where \(x_{i-1/2} = (i - 1/2)h_x\), \(y_{j-1/2} = (j - 1/2)h_y\), and \(z_{\ell-1/2} = (\ell - 1/2)h_z\) define cell-centered grid positions. The index range \(-1 \leq i,j,\ell \leq N+2\) is adopted to accommodate boundary extrapolation.
To enforce the homogeneous Neumann boundary condition \cref{boundary-large} while preserving high-order accuracy, a third-order extrapolation scheme is implemented. For instance, along the \(z\)-direction (normal to the boundary at \(z=0\) and \(z=L\)), the extrapolation rules for the magnetization field are:
\begin{align*}
\m_{i,j,1}=\m_{i,j,0},\quad \m_{i,j,-1}=\m_{i,j,2},\quad \m_{i,j,N+1}=\m_{i,j,N} ,\quad \m_{i,j,N+2}=\m_{i,j,N-1}.
\end{align*}
Analogous extrapolation formulas are applied to enforce the boundary condition along the \(x\)- and \(y\)-directions, ensuring consistent high-order accuracy across the entire computational domain.

The fourth-order spatial difference operators should be introduced further. 
To achieve fourth-order spatial accuracy, which is essential for resolving fine magnetic structures (e.g., domain walls with nanoscale width) and minimizing discretization-induced errors, we employ long-stencil finite difference operators for first and second partial derivatives. For the \(x\)-direction, the fourth-order accurate operators for \(\partial_x\) (denoted \({\mathcal D}_{x,(4)}^1\)) and \(\partial_x^2\) (denoted \({\mathcal D}_{x,(4)}^2\)) are defined as:
\begin{eqnarray} 
\hspace{-0.35in}  
{\mathcal D}_{x,(4)}^1 f_{i,j,k} &=& \tilde{D}_x ( 1 - \frac{h_x^2}{6} D_x^2 ) f_{i,j,k} \nonumber 
\\
&=& 
\frac{  f_{i-2,j,k} - 8 f_{i-1,j,k}  + 8 f_{i+1,j,k} - f_{i+2,j,k} }{12 h_x} ,  
\label{FD-4th-1} 
\\
\hspace{-0.35in}  
{\mathcal D}_{x,(4)}^2 f_{i,j,k} &=& D_x^2 ( 1 - \frac{h_x^2}{12} D_x^2 ) f_{i,j,k}  \nonumber 
\\
&=& 
\frac{ - f_{i-2,j,k} + 16 f_{i-1,j,k,k} - 30 f_{i,j,k} + 16 f_{i+1,j,k} - f_{i+2,j,k} }{12 h_x^2 } . 
\label{FD-4th-2} 
\end{eqnarray} 
These operators are derived by canceling leading-order discretization errors via inclusion of adjacent grid points, resulting in fourth-order convergence (\(O(h^4)\)) for sufficiently smooth functions.

By symmetry, the fourth-order difference operators for the \(y\)- and \(z\)-directions—designated \({\mathcal D}_{y,(4)}^1\), \({\mathcal D}_{y,(4)}^2\), \({\mathcal D}_{z,(4)}^1\), and \({\mathcal D}_{z,(4)}^2\)—are defined by substituting the respective spatial coordinates and mesh-sizes. The discrete Laplacian operator \(\Delta_h\), which approximates the continuous Laplacian \(\Delta\) with fourth-order accuracy, is constructed as the sum of the second-order operators in all three directions: \(\Delta_h = {\mathcal D}_{x,(4)}^2 + {\mathcal D}_{y,(4)}^2 + {\mathcal D}_{z,(4)}^2\).

To provide a rigorous performance baseline for the proposed numerical method, we adapt the first-order (BDF1) and second-order (BDF2) backward differentiation formula schemes, which are standard in micromagnetics, to include the fourth-order spatial operators and the equivalent LLG formulation \cref{eq-model}. To better understand the process of stability, the proposed method is compared with an existing third order numerical scheme. These adapted BDF schemes will serve as benchmarks in subsequent numerical experiments, allowing quantitative evaluation of the proposed method's accuracy, efficiency, and stability.

\subsection{The first order semi-implicit method}

The BDF1 scheme is as follows,
\begin{equation}\label{BDF1}
\left\{ 
\begin{aligned}
&\frac{{\tilde{\m}}_h^{n+1} -  {\m}_h^n}{k}
=  - {\m}_h^{n} \times\big(\epsilon \Delta_h\tilde{\m}_h^{n+1} +{\f}_h^{n} \big) \\
&\quad - \alpha {\m}_h^{n} \times \left({\m}_h^{n}\times(\epsilon \Delta_h\tilde{\m}_h^{n+1} +{\f}_h^{n} ) \right), \\
& \qquad\qquad\qquad\qquad\quad \m_h^{n+1} = \frac{\tilde{\m}_h^{n+1}}{ |\tilde{\m}_h^{n+1}| },
\end{aligned}
\right.
\end{equation} 
The cross product terms in BDF1 method lead to a linear system with non-symmetric structure and variable coefficients. Consequently, the GMRES solver is typically employed to solve this numerical system.

\subsection{The second order semi-implicit method}
The second-order semi-implicit projection method (SIPM) has been described in \cite{Xie2018,jingrun2019analysis}. This approach is based on a second-order BDF temporal discretization combined with explicit extrapolation. It has been established that SIPM is unconditionally stable and achieves second-order accuracy in both space and time. The algorithm is detailed below.
\begin{equation}\label{sipm}
\left\{ 
\begin{aligned}
&\frac{\frac32 {\tilde{\m}}_h^{n+2} - 2 {\m}_h^{n+1} + \frac12 {\m}_h^n}{k}
=  - \hat{\m}_h^{n+2} \times\big(\epsilon \Delta_h\tilde{\m}_h^{n+2} +\hat{\f}_h^{n+2} \big) \\
&\quad - \alpha \hat{\m}_h^{n+2} \times \left(\hat{\m}_h^{n+2}\times(\epsilon \Delta_h\tilde{\m}_h^{n+2} +\hat{\f}_h^{n+2} ) \right), \\
& \qquad\qquad\qquad\qquad\quad \m_h^{n+2} = \frac{\tilde{\m}_h^{n+2}}{ |\tilde{\m}_h^{n+2}| },
\end{aligned}
\right.
\end{equation} 
where $\tilde{\m}_h^{n+2}$ is an intermediate magnetization, and $\hat{\m}_h^{n+2}$, $\hat{\f}_h^{n+2}$ are given by the extrapolation formula: 
\begin{align*}
\hat{\m}_h^{n+2} &=2{\m}_h^{n+1}-{\m}_h^n, \label{m_hat}\\
\hat{\f}_h^{n+2} &=2{\f}_h^{n+1}-{\f}_h^n,
\end{align*}
with $\f_h^{n}=-Q(m_2^n\e_2+m_3^n\e_3)+\h_s^n+\h_e^n$.

\subsection{The third order semi-implicit method}
A third-order semi-implicit scheme for model \cref{eq-model} as presented in \cite{xie2025thirdorder}, is given by
\begin{equation}\label{scheme-third-order}
\left\{ 
\begin{aligned}
&\frac{\frac{11}{6}\tilde{\m}_h^{n+3}-3{\m}_h^{n+2}+\frac32{\m}_h^{n+1}-\frac13 {\m}_h^n}{k}\\
&\quad=-\hat{\m}_h^{n+3} \times (\epsilon \Delta_{h,(4)}\tilde{\m}_h^{n+3}+\hat{\f}_h^{n+3})+\alpha(\epsilon \Delta_{h,(4)}\tilde{\m}_h^{n+3} +\hat{\f}_h^{n+3})\\
&\qquad+\alpha(\epsilon |\tilde{\nabla}_{h,(4)}\hat{\m}_h^{n+3}|^2+\hat{\m}_h^{n+3}\cdot \hat{\f}_h^{n+3}) \hat{\m}_h^{n+3}, \\
&\m_h^{n+3}=\frac{\tilde{\m}_h^{n+3}}{|\tilde{\m}_h^{n+3}|},
\end{aligned}
\right.
\end{equation}
where
\begin{align*}
\hat{\m}_h^{n+3} &= 3 \m_h^{n+2}-3\m_h^{n+1} + \m_h^n,\\
\hat{\f}_h^{n+3} &= 3 \f_h^{n+2}-3\f_h^{n+1} + \f_h^n.
\end{align*}

\subsection{The proposed numerical method} \label{discretisations}

The SIPM in \eqref{sipm} employs a semi-implicit treatment for both the gyromagnetic and damping terms, meaning $\Delta \m$ is handled implicitly while coefficient functions are updated via a second-order explicit extrapolation. A natural extension is to apply a third-order BDF method, where $\Delta \m$ remains implicit and coefficient functions are extrapolated using a third-order accurate explicit formula. This concept leads to the proposed method:

\begin{equation}\label{proposed}
\left\{ 
\begin{aligned}
&\frac{\frac{11}{6} \tilde{\m}_h^{n+3} - 3 {\m}_h^{n+2} + \frac32 {\m}_h^{n+1}-\frac13 {\m}_h^n}{k}
=  - \hat{\m}_h^{n+3} \times \left(\epsilon \Delta_{h,(4)} \tilde{{\m}}_h^{n+3} +\hat{\f}_h^{n+3}\right) \\
&\quad -\alpha\hat{\m}_h^{n+3}\times (\hat{\m}_h^{n+3}\times (\epsilon\Delta_{h,(4)}\tilde{{\m}}_h^{n+3}+\hat{\f}_h^{n+3}))\\ 
&\m_h^{n+3} = \frac{\tilde{\m}_h^{n+3}}{ |\tilde{\m}_h^{n+3}| } ,
\end{aligned}
\right.
\end{equation}
where
\begin{align*}
\hat{\m}_h^{n+3} &= 3 \m_h^{n+2} - 3\m_h^{n+1}+\m_h^{n},\\
\hat{\f}_h^{n+3} &= 3 \f_h^{n+2} - 3\f_h^{n+1}+{\f}_h^n.
\end{align*}

In terms of formal performance, the proposed method demonstrates distinct advantages over both BDF1 method and BDF2 method, especially regarding accuracy and computational efficiency. As detailed in the numerical results of \cref{sec:experiments}, the proposed scheme proves to be a reliable and robust tool for micromagnetic simulations, offering high accuracy and efficiency particularly for moderate damping parameters. The stability advantage of the proposed method over the third-order semi-implicit scheme \cref{scheme-third-order} is also demonstrated in subsequent simulations.

\begin{remark}
Mathematically, while \cref{eq-5} and \cref{eq-model} are equivalent under the normalization condition $|\m|=1$, their numerical implementations in schemes \cref{scheme-third-order} and \cref{proposed}  yield different performances. The differences between the schemes are as follows.
\begin{itemize}
	\item The damping terms differ because $|\hat{\m}^{n+3}|\neq 1$ and $\hat{\m}^{m+3}\cdot\Delta \tilde{{\m}}_h^{n+3}\neq |\nabla \hat{\m}^{n+3}|$;
	\item The coefficient matrices of the resulting linear systems differ: scheme \cref{scheme-third-order} yields $\frac{11}{6}I+\epsilon \hat{\m}^{n+3}\times \Delta_h+\alpha \epsilon \Delta_h$, whereas scheme \cref{proposed} yields $\frac{11}{6}I+\epsilon \hat{\m}^{n+3}\times \Delta_h+\alpha \hat{\m}^{n+3}\times (\hat{\m}^{n+3}\times \Delta_h)$. The right-hand side source terms also differ.
\end{itemize}
\end{remark}

\section{Numerical experiments}
\label{sec:experiments}

This section provides a systematic numerical evaluation of four micromagnetic methods: the proposed third-order scheme, the second-order BDF2, and the first-order BDF1, the existing third order semi-implicit scheme. To encompass practical scenarios, experiments employ a designated range of Gilbert damping parameters, covering weak, moderate, and strong dissipation regimes. Performance is assessed using three key metrics: accuracy, efficiency, and stability, with detailed data provided for direct comparison. Accuracy is verified using exact solutions and convergence rates; efficiency via computational time and error; stability by monitoring magnetization profiles and energy dissipation. Special attention is given to domain wall dynamics, which are critical for magnetic devices. Domain wall velocity is consistent with previous work. This validates methodological consistency and elucidates the interplay between damping, field, and velocity, aiding magnetic device optimization.

\subsection{Accuracy and efficiency tests}

To simplify temporal accuracy analysis, we set parameter  \(\epsilon=1\) and forcing term \(\f=0\) in the governing model \cref{eq-model}. Analytical exact solutions are derived for both one-dimensional (1D) and three-dimensional (3D) scenarios to serve as benchmarks for error quantification.

For the 1D case, the exact magnetization solution \(\m_e\) is:
\begin{equation*}
\m_e=\left(\cos(\cos(\pi x))\sin t, \sin(\cos(\pi x))\sin t, \cos t\right)^T,
\end{equation*}
while the corresponding 3D exact solution is:
\begin{equation*}
\m_e=\left(\cos(\cos(\pi x)\cos(\pi y)\cos(\pi z))\sin t, \sin(\cos(\pi x)\cos(\pi y)\cos(\pi z))\sin t, \cos t\right)^T,
\end{equation*}
These exact solutions satisfy the governing equation \cref{eq-model} when the forcing term is defined as \(\g=\partial_t \m_e-\alpha \Delta \m_e -\alpha |\nabla \m_e|^2+\m_e \times \Delta \m_e\). They also comply with the homogeneous Neumann boundary condition, ensuring consistency with simulation constraints.

To isolate the temporal approximation error from spatial discretization effects, the spatial resolution in the 1D test is fixed at \(h=10^{-4}\)—a sufficiently fine grid that makes spatial error negligible compared to temporal error. The Gilbert damping parameter is set to \(\alpha=0.01\), and simulations run until final time \(T=0.1\). Under this configuration, the measured error primarily reflects temporal discretization inaccuracy.

The 3D temporal accuracy test faces inherent constraints from spatial resolution, as excessively fine grids incur prohibitive computational cost. To balance spatial and temporal error contributions, we adopt a coordinated refinement strategy for spatial mesh sizes (\(h_x, h_y, h_z\)) and temporal step-size (\(k\)) tailored to each method's order:
\begin{itemize}
	\item For first-order BDF1: \(k=h_x^2=h_y^2=h_z^2=h^2=T/N_0\);
	\item For second-order BDF2: \(k=h_x=h_y=h_z=h=T/N_0\);
	\item For the proposed third-order method: \(k=h_x^{4/3}=h_y^{4/3}=h_z^{4/3}=h^{4/3}=T/N_0\).
\end{itemize}
Here, \(N_0\) is a refinement level parameter, with specific values given in subsequent results. Consistent with the 1D test, \(\alpha=0.01\), and the final time \(T\) is specified later.

Numerical errors for the three methods (BDF1, BDF2, proposed) are systematically recorded versus temporal step size \(k\), with detailed data compiled in \cref{tab-1}. Error analysis clearly identifies the temporal accuracy orders: the proposed method achieves third-order (\(3\)) accuracy, while BDF1 and BDF2 exhibit first-order (\(1\)) and second-order (\(2\)) accuracy, respectively. This order consistency holds for both 1D and 3D scenarios, confirming the method's reliability across spatial dimensions.

\begin{table}[htbp] 
	\centering
	{\caption{The numerical errors for the proposed method, the BDF1 and the BDF2 with $\alpha=0.01$ and $T=0.1$. Left: 1D with $h=1D-4$; Right: 3D with $k=h_x^2=h_y^2=h_z^2=h^2=T/N_0$ for BDF1 and $k=h_x=h_y=h_z=h=T/N_0$ for the BDF2 method and and $k=h_x^{4/3}=h_y^{4/3}=h_z^{4/3}=h^{4/3}=T/N_0$ for the proposed method, with $N_0$ specified in the table.}\label{tab-1} }{
		\subfloat[Proposed method]{\label{tab:floatrow:one}%
			\begin{tabular}{cccc|cccc} 
				\hline	
				1D  & & {} & {} &3D &{} & {} &{} \\
				$k$ & $\|\cdot\|_{\infty}$ & $\|\cdot\|_{2}$ & $\|\cdot\|_{H^1}$ & 	$k, k^3\approx h^4$ & $\|\cdot\|_{\infty}$ & $\|\cdot\|_{2}$ & $\|\cdot\|_{H^1}$ \\
				\hline
			$T/8$& 3.525D-9& 2.932D-9 & 9.275D-9& $T/6$ & 3.111D-6 & 5.690D-7 & 1.000D-5 \\	
			$T/12$ & 1.156D-9& 9.134D-10 & 3.295D-9 & $T/7$  & 1.950D-6 & 3.889D-7 & 7.231D-6\\
			$T/16$ & 5.085D-10 & 3.960D-10 & 1.513D-9&	$T/8$ & 1.043D-6 & 2.362D-7 & 4.579D-6\\
			$T/24$ & 1.552D-10 & 1.201D-10 & 4.800D-10 &	 $T/9$ & 7.359D-7 & 1.748D-7 & 3.458D-6\\
			$T/32$ & 6.664D-11 & 5.135D-11 & 2.087D-10 & 	$T/11$ & 4.429D-7 & 9.233D-8 & 1.828D-6\\
				order &2.87 &2.92& 2.74&{--}&3.30&3.03&2.83\\
				
				
				%
				\hline
			\end{tabular}	
		}
		\qquad
		\subfloat[BDF1]{\label{tab:floatrow:two}
			\begin{tabular}{cccc|cccc} 
				\hline
				1D & &  & {} & 3D & {} &  & {} \\
				$k$ & $\|\cdot\|_{\infty}$ & $\|\cdot\|_{2}$ & $\|\cdot\|_{H^1}$ & $k=h^2$ & $\|\cdot\|_{\infty}$ & $\|\cdot\|_{2}$ & $\|\cdot\|_{H^1}$\\
				\hline
				$T/8$ & 2.574D-5 & 2.341D-5 & 7.999D-5& $T/40$ & 4.147D-4 & 6.722D-5 & 5.380D-4 \\
				$T/12$ &1.776D-5 & 1.517D-5 & 5.591D-5 & $T/57$ & 2.900D-4 & 4.682D-5 & 3.729D-4 \\
				$T/16$ &1.358D-5 & 1.126D-5 & 4.324D-5 & $T/78$ & 2.139D-4 & 3.446D-5 & 2.736D-4  \\
				$T/24$ & 9.240D-6 & 7.444D-6 & 2.982D-5 &$T/102$ &1.642D-4 & 2.642D-5 & 2.093D-4 \\
				$T/32$ & 7.003D-6 & 5.566D-6 & 2.276D-5 &$T/129$ & 1.300D-4 & 2.089D-5 & 1.652D-4\\
				order & 0.94&1.03&0.91 &{--}&0.99 &1.00&1.01 \\
				
				
				%
				\hline
			\end{tabular}
		}
		\qquad
		\subfloat[BDF2]{\label{tab:floatrow:three}
			\begin{tabular}{cccc|cccc} 
				\hline
				1D  & & & {} & 3D & {} &  & {}\\
				$k$ & $\|\cdot\|_{\infty}$ & $\|\cdot\|_{2}$ & $\|\cdot\|_{H^1}$ & $k=h$ & $\|\cdot\|_{\infty}$ & $\|\cdot\|_{2}$ & $\|\cdot\|_{H^1}$\\
				\hline
				$T/8$ & 7.975D-6 & 6.714D-6 & 3.160D-5& $T/2$ &2.291D-4 & 3.243D-5 & 7.541D-4\\	
			$T/12$ & 3.661D-6 & 3.067D-6 & 1.418D-5& $T/3$ & 1.153D-4 & 1.890D-5 & 3.258D-4 \\
				$T/16$ & 2.085D-6 & 1.745D-6 & 7.999D-6 &	$T/4$ &6.530D-5 & 1.178D-5 & 1.824D-4\\
			$T/24$ &9.364D-7 & 7.836D-7 & 3.561D-6&	 $T/5$ &4.356D-5 & 7.950D-6 & 1.168D-4 \\
				$T/32$ & 5.288D-7 & 4.428D-7 & 2.005D-6  & $T/6$& 3.090D-5 & 5.699D-6 & 8.128D-5 \\
				order &1.96&1.96&1.99 & {--}&1.83 &1.59&2.03\\
				
				
				%
				\hline
			\end{tabular}
	} }	
\end{table}

Following temporal accuracy evaluation, spatial accuracy tests were conducted to quantify the spatial discretization performance of the proposed third-order method, BDF1, and BDF2. To prevent temporal errors from interfering with spatial accuracy assessment, the temporal step size was fixed at a sufficiently small \(k=10^{-5}\) for all 1D and 3D tests—making temporal errors negligible compared to spatial errors. Consistent with the temporal test, \(\alpha=0.01\), and all simulations ran until \(T=0.1\). The exact solutions described earlier (for 1D and 3D) were used as benchmarks to compute numerical errors. Specifically, the \(L^2\)-norm error between the numerical and exact solutions was computed for a sequence of decreasing spatial grid sizes \(h\), isolating the spatial discretization impact. Spatial error data for all three methods, organized by \(h\), are systematically documented in \cref{tab-2}. A comprehensive analysis reveals distinct spatial accuracy characteristics: BDF1 and BDF2 both achieve second-order spatial accuracy, while the proposed method demonstrates higher fourth-order spatial accuracy. This disparity in spatial accuracy orders is consistent across 1D and 3D scenarios, further validating the superior spatial discretization capability of the proposed method.

\begin{table}[htbp]
	\centering
	{\caption{The numerical errors of the proposed method, the BDF1 and the BDF2 with $\alpha=0.01$ and $T=0.1$. Left: 1D with $k=1D-6$; Right: 3D with $k=1D-5$.} \label{tab-2} }{
		\subfloat[Proposed method]{\label{tab:floatrow:1-S}
			\begin{tabular}{cccc|cccc}	
				\hline
				1D  & & & {} & 3D & & & \\
				$h$ & $\|\cdot\|_{\infty}$ &$\|\cdot\|_{2}$ &$\|\cdot\|_{H^1}$& $h$ & $\|\cdot\|_{\infty}$ & $\|\cdot\|_{2}$ & $\|\cdot\|_{H^1}$  \\
				\hline
				1/16 & 9.094D-6 & 6.487D-6 & 9.920D-5 & 1/4 & 1.466D-3 & 3.971D-4 & 6.935D-3\\
				1/24 &1.829D-6 & 1.296D-6 & 2.004D-5& 1/6 & 3.509D-4 & 8.772D-5 & 1.574D-3\\
				1/32 & 5.811D-7 & 4.117D-7 & 6.394D-6 & 1/8 & 1.267D-4 & 2.897D-5 & 5.305D-4\\
				1/48 & 1.149D-7 & 8.157D-8 & 1.270D-6& 1/10 & 5.825D-5 & 1.213D-5 & 2.244D-4 \\
				1/64 & 3.647D-8 & 2.583D-8 & 4.028D-7& 1/12 & 2.987D-5 & 5.928D-6 & 1.101D-4 \\
				order  & 3.98&3.99&3.97 & {--} &3.54 &3.83&3.77  \\
				\hline
			\end{tabular}
		}	
		\qquad
		\subfloat[BDF1]{\label{tab:floatrow:2-S}
			\begin{tabular}{cccc|cccc}	
				\hline
				1D  & &  & {} & 3D & & & \\
				$h$ & $\|\cdot\|_{\infty}$ &$\|\cdot\|_{2}$ &$\|\cdot\|_{H^1}$ & $h$ & $\|\cdot\|_{\infty}$ &$\|\cdot\|_{2}$ &$\|\cdot\|_{H^1}$\\
				\hline
				1/16 & 4.223D-4 & 2.893D-4 & 2.216D-3 & 1/4 & 8.026D-3 & 1.816D-3 & 1.589D-2\\
				1/24 & 1.883D-4 & 1.283D-4 & 9.801D-4& 1/6 & 4.094D-3 & 7.779D-4 & 6.575D-3\\
				1/32 & 1.060D-4 & 7.208D-5 & 5.503D-4 & 1/8 & 2.437D-3 & 4.326D-4 & 3.570D-3\\
				1/48 & 4.717D-5 & 3.201D-5 & 2.443D-4& 1/10 & 1.611D-3 & 2.753D-4 & 2.242D-3\\
				1/64 & 2.654D-5 & 1.800D-5 & 1.373D-4& 1/12 & 1.139D-3 & 1.906D-4 & 1.539D-3 \\
				order  & 2.00&2.00&2.01 & {--} & 1.78&2.05& 2.13\\
				\hline
			\end{tabular}
		}
		\qquad
		\subfloat[BDF2]{\label{tab:floatrow:3-S}
			\begin{tabular}{cccc|cccc}	
				\hline
				1D  & &  & & 3D & & & \\
				$h$ & $\|\cdot\|_{\infty}$ &$\|\cdot\|_{2}$ &$\|\cdot\|_{H^1}$ &$h$ & $\|\cdot\|_{\infty}$ &$\|\cdot\|_{2}$ &$\|\cdot\|_{H^1}$ \\
				\hline
				1/16 & 4.223D-4 & 2.894D-4 & 2.216D-3 & 1/4 & 8.027D-3 & 1.817D-3 & 1.589D-2\\
				1/24 & 1.883D-4 & 1.283D-4 & 9.801D-4& 1/6 & 4.094D-3 & 7.779D-4 & 6.576D-3\\
				1/32 & 1.060D-4 & 7.208D-5 & 5.503D-4 & 1/8 & 2.437D-3 & 4.326D-4 & 3.570D-3\\
				1/48 & 4.717D-5 & 3.202D-5 & 2.443D-4& 1/10 & 1.611D-3 & 2.753D-4 & 2.242D-3 \\
				1/64 & 2.654D-5 & 1.800D-5 & 1.373D-4& 1/12 & 1.139D-3 & 1.906D-4 & 1.539D-3\\
				order  &2.00&2.00&2.01& {--} &1.78&2.05&2.13 \\
				\hline
			\end{tabular} 
		}
	}
\end{table}

\begin{figure}[htbp]
	\centering
	\subfloat[Varying $k$ in 1D up to $T=0.1$ ]{\label{cputime_1D}\includegraphics[width=2.5in]{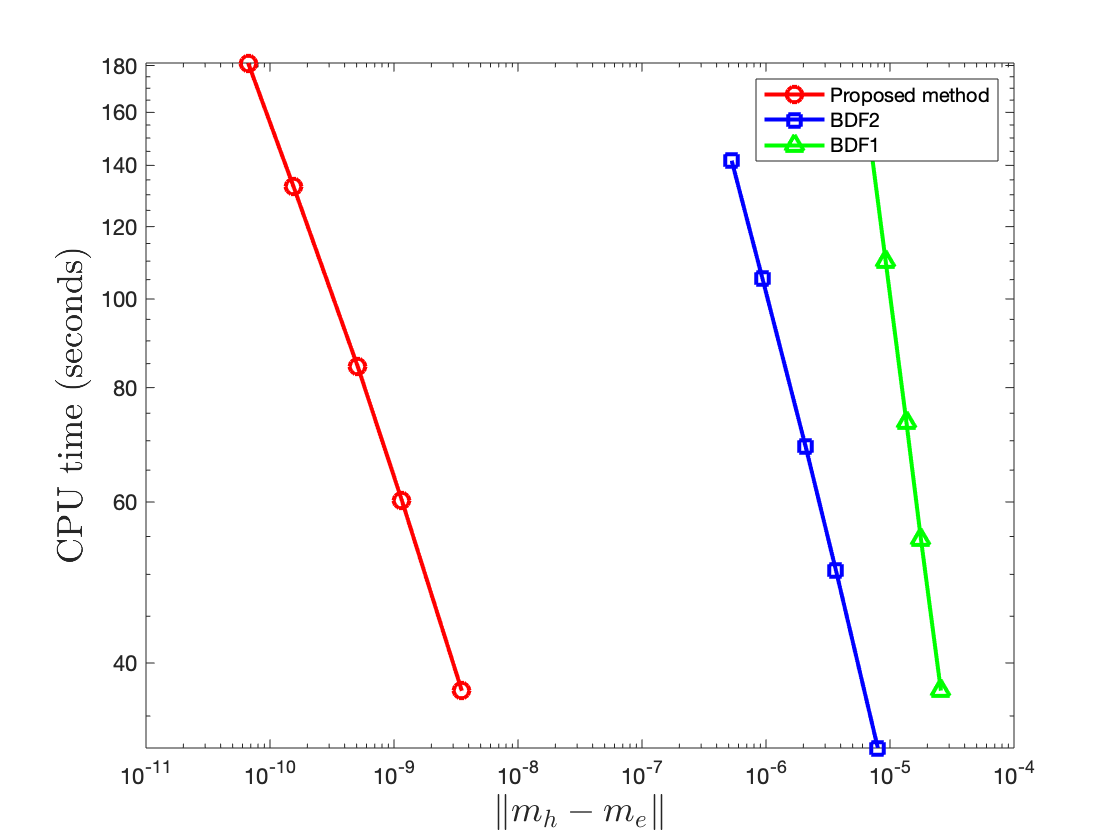}}
	\subfloat[Varying $k$ in 3D up to $T=0.1$]{\label{cputime_3D}\includegraphics[width=2.5in]{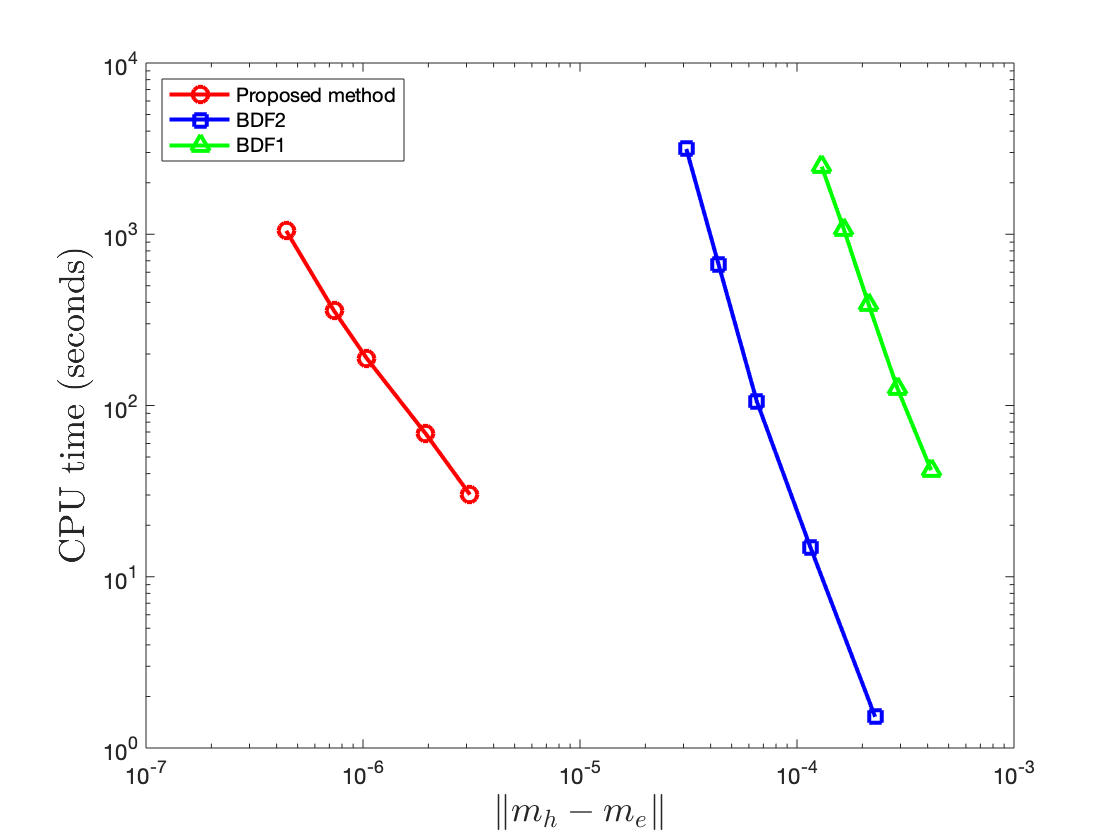}}
	\hspace{0.1in}
	\subfloat[Varying $h$ in 1D up to $T=0.1$]{\label{cputime_1D_space}\includegraphics[width=2.5in]{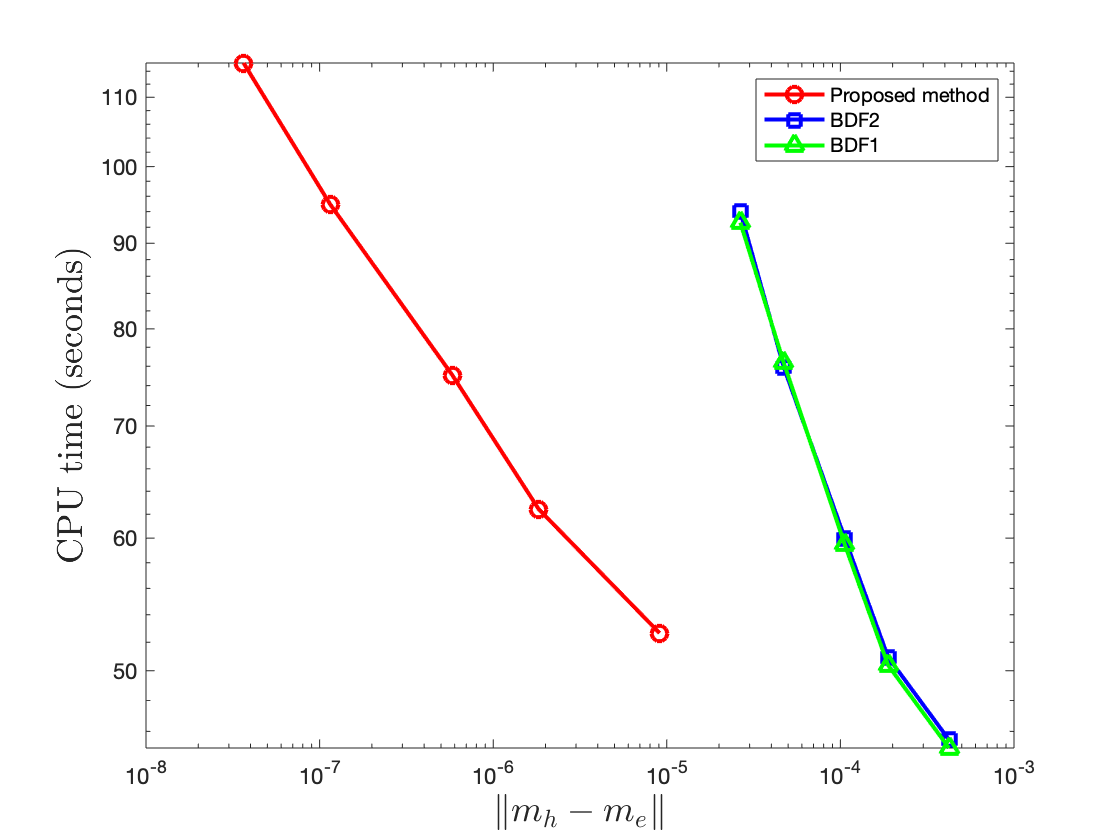}}
	\subfloat[Varying $h$ in 3D up to $T=0.1$]{\label{cputime_3D_space}\includegraphics[width=2.5in]{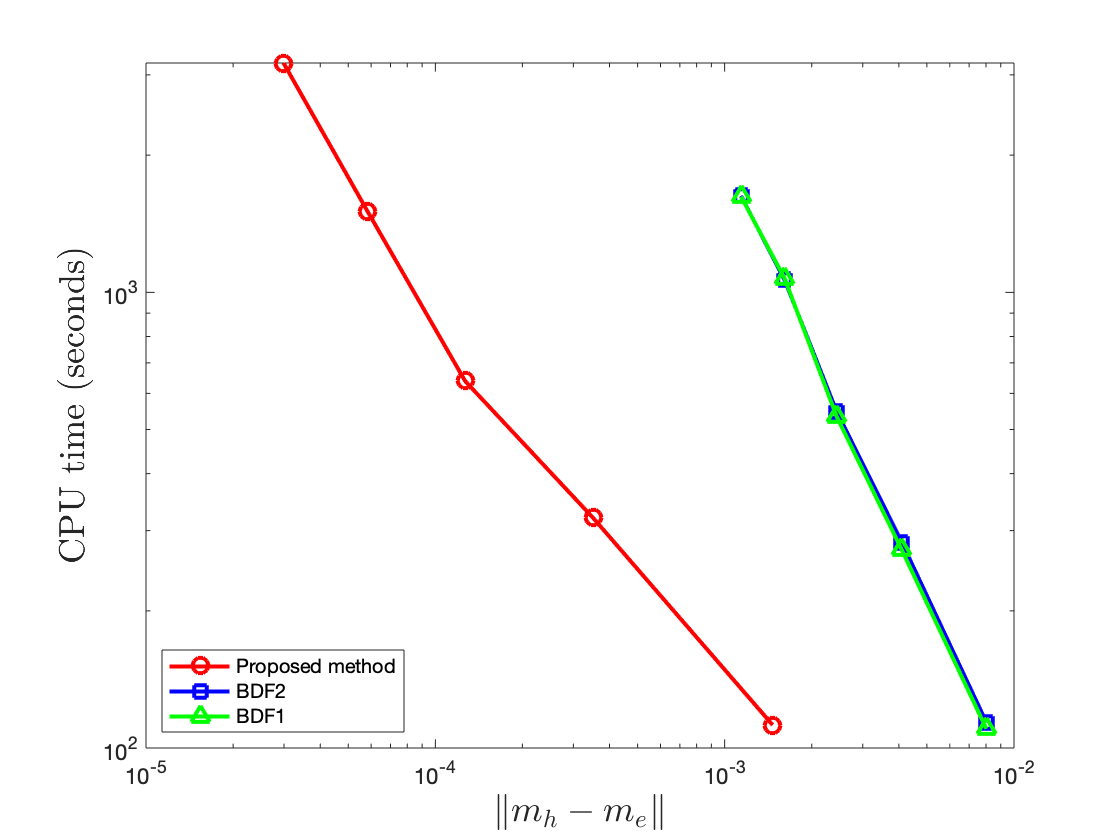}}
	\caption{CPU time required to achieve the desired numerical accuracy for the proposed method, BDF2, and BDF1 in both 1D and 3D computations. CPU time is recorded as a function of approximation error by varying $k$ or $h$ independently. CPU time with varying $k$: proposed method $<$ BDF2 $<$ BDF1; CPU time with varying $h$: proposed method $<$ BDF1 $\lessapprox$ BDF2.}\label{cputime}
\end{figure}

To compare numerical efficiency, we plot CPU time (in seconds) versus the error norm $\|\m_h-\m_e\|_{\infty}$. Specifically, CPU time as a function of approximation error is shown in \cref{cputime_1D} for 1D and in \cref{cputime_3D} for 3D, with varying $k$ and fixed $h$. Similar plots are displayed in \cref{cputime_1D_space} for 1D and \cref{cputime_3D_space} for 3D, with varying $h$ and fixed $k$. For fixed spatial resolution $h$, the proposed method is significantly more efficient than BDF1 and BDF2 in both 1D and 3D computations. BDF2 is slightly more efficient than BDF1, though this advantage may vary with $k$ and $h$.For fixed time step size $k$, the proposed method is slightly more efficient than the BDF2 and BDF1, in both the 1D and 3D computations, with BDF1 cost being comparable to BDF2.

\begin{figure}[htbp]
	\centering
	\subfloat[Varying $k$ (1D) ]{\label{cputime_ED_1D_time}\includegraphics[width=1.5in]{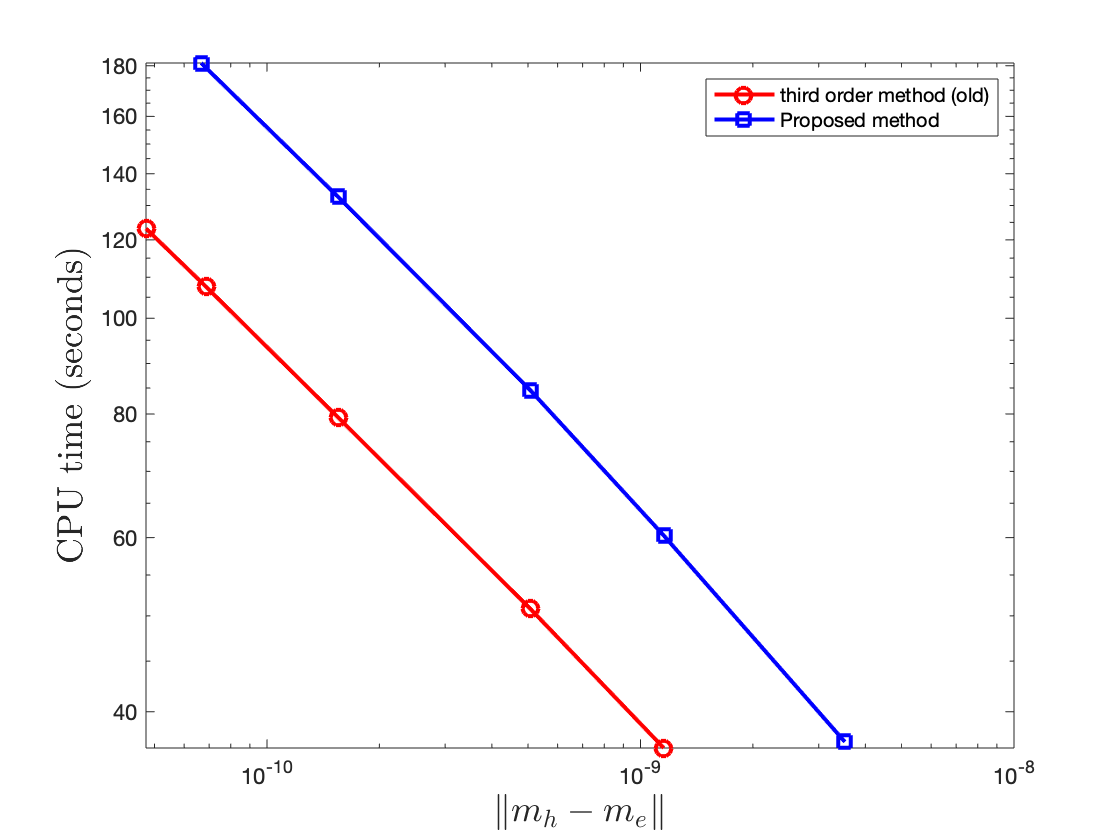}}
		\subfloat[Varying $k$ (3D) ]{\label{cputime_ED_3D_time}\includegraphics[width=1.5in]{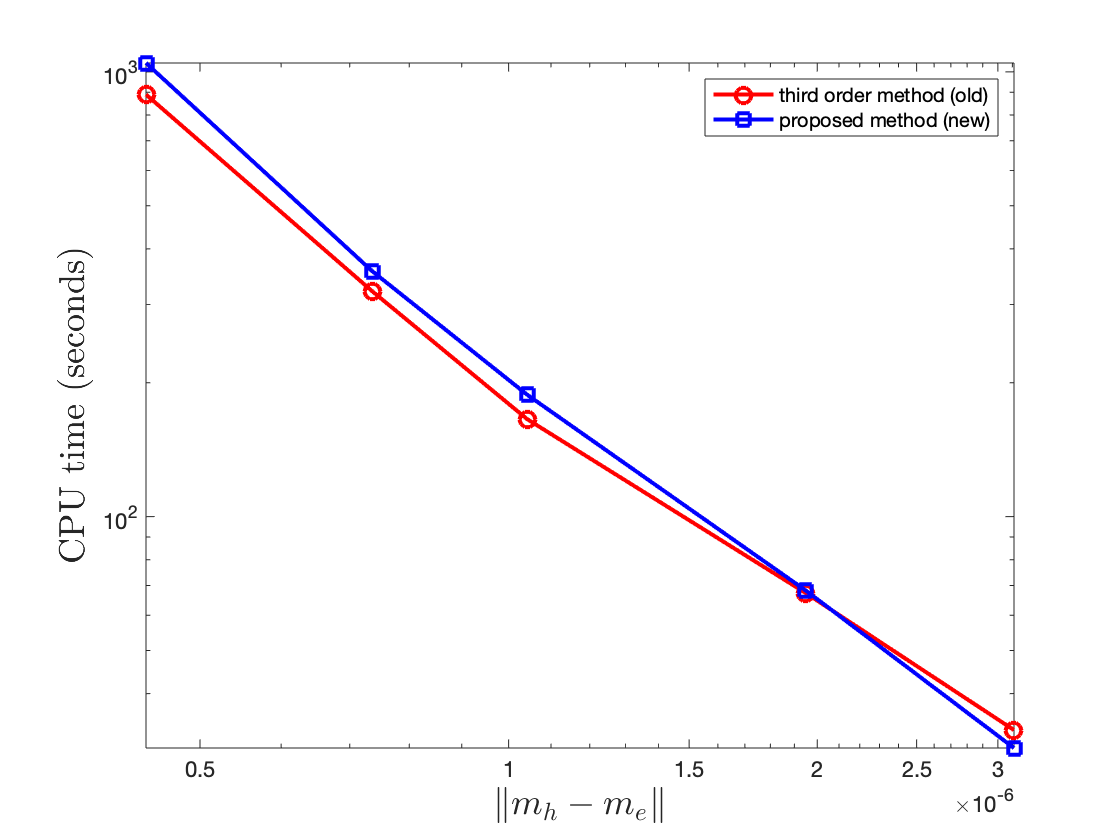}}
	\subfloat[Varying $h$ (1D) ]{\label{cputime_ED_1D_space}\includegraphics[width=1.5in]{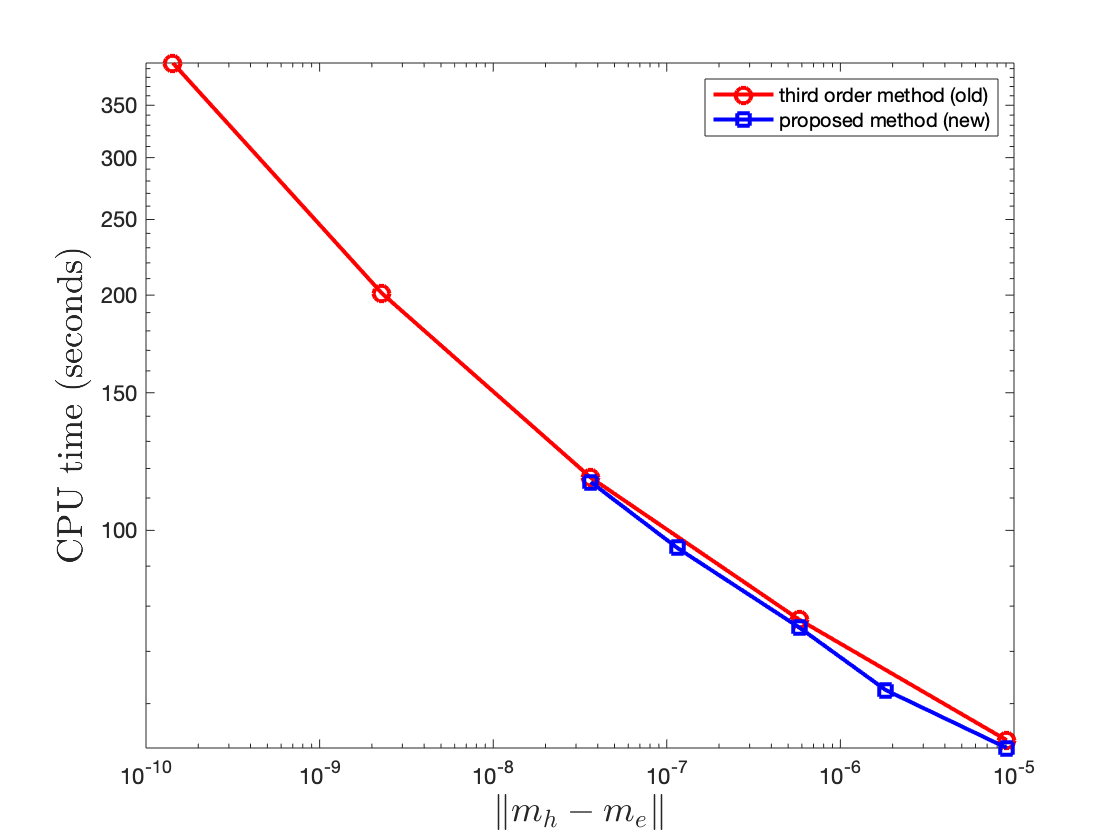}}
		\subfloat[Varying $h$ (3D) ]{\label{cputime_ED_3D_space}\includegraphics[width=1.5in]{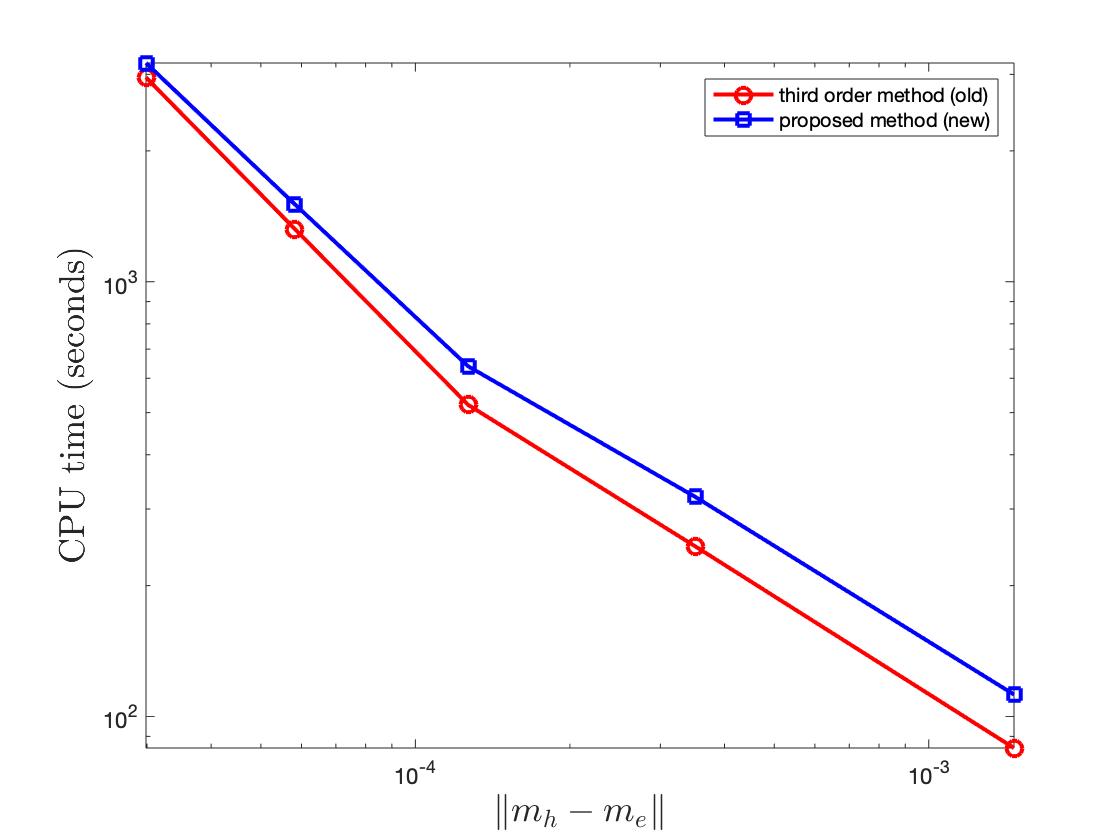}}
		\hspace{0.1in}
		\subfloat[Varying $k$ (1D) ]{\label{cputime_ED_BDF2_1D_time}\includegraphics[width=1.5in]{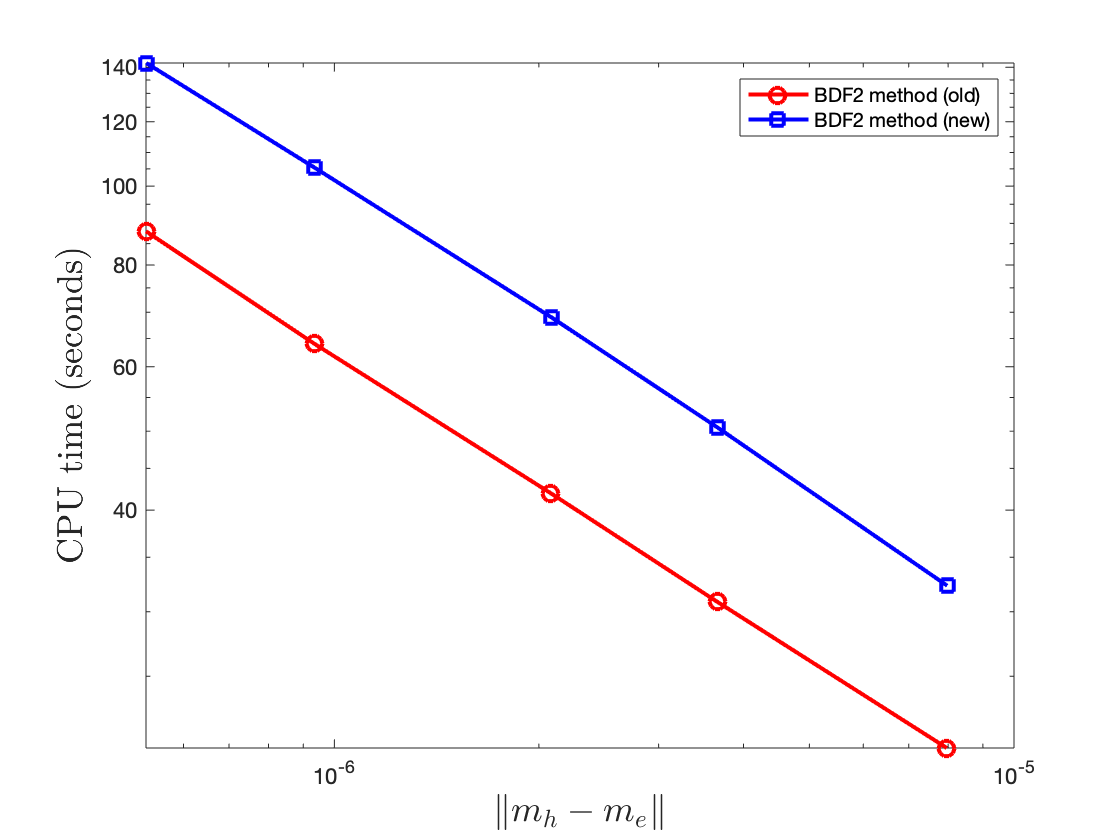}}
		\subfloat[Varying $k$ (3D) ]{\label{cputime_ED_BDF2_3D_time}\includegraphics[width=1.5in]{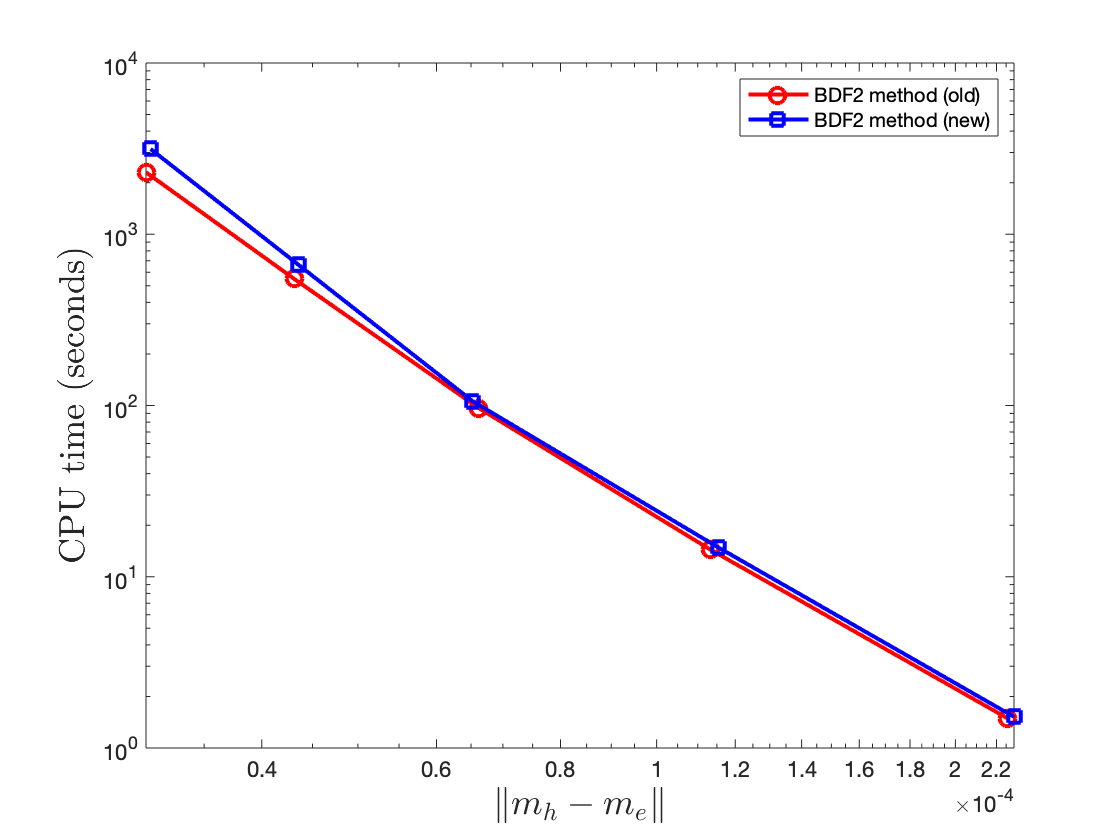}}
		\subfloat[Varying $h$ (1D) ]{\label{cputime_ED_BDF2_1D_space}\includegraphics[width=1.5in]{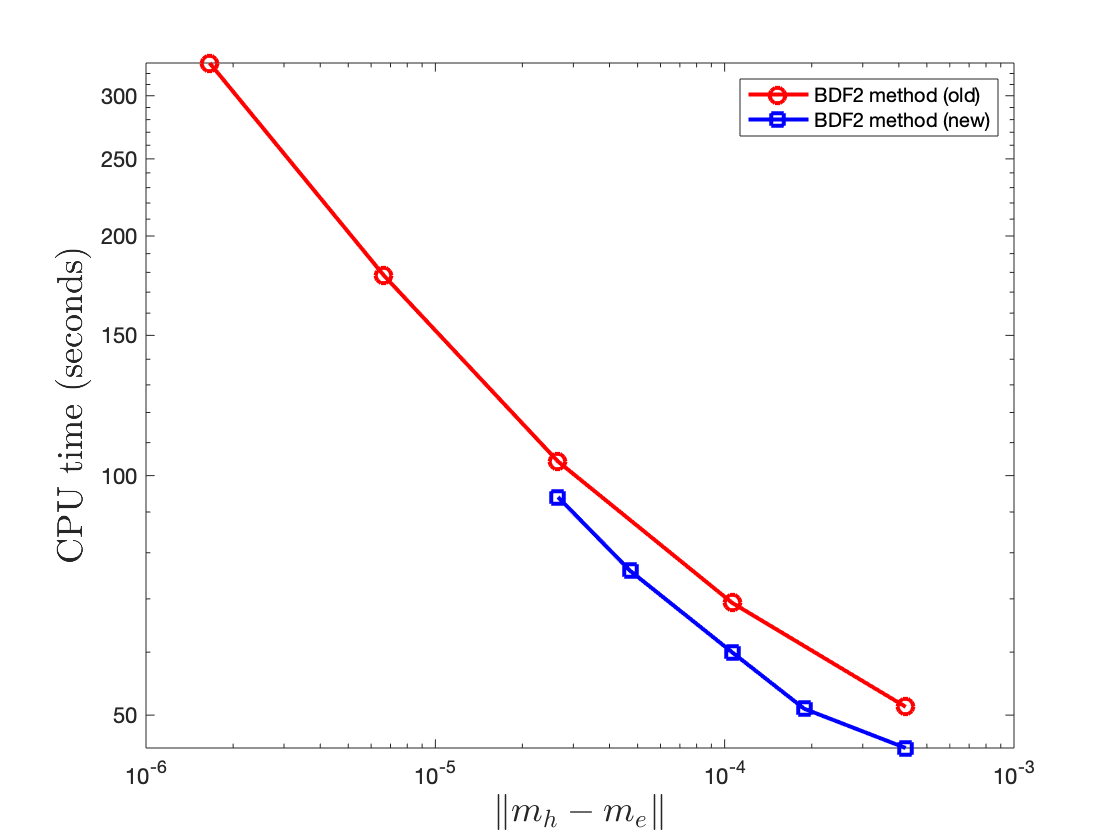}}
		\subfloat[Varying $h$ (3D) ]{\label{cputime_ED_BDF2_3D_space}\includegraphics[width=1.5in]{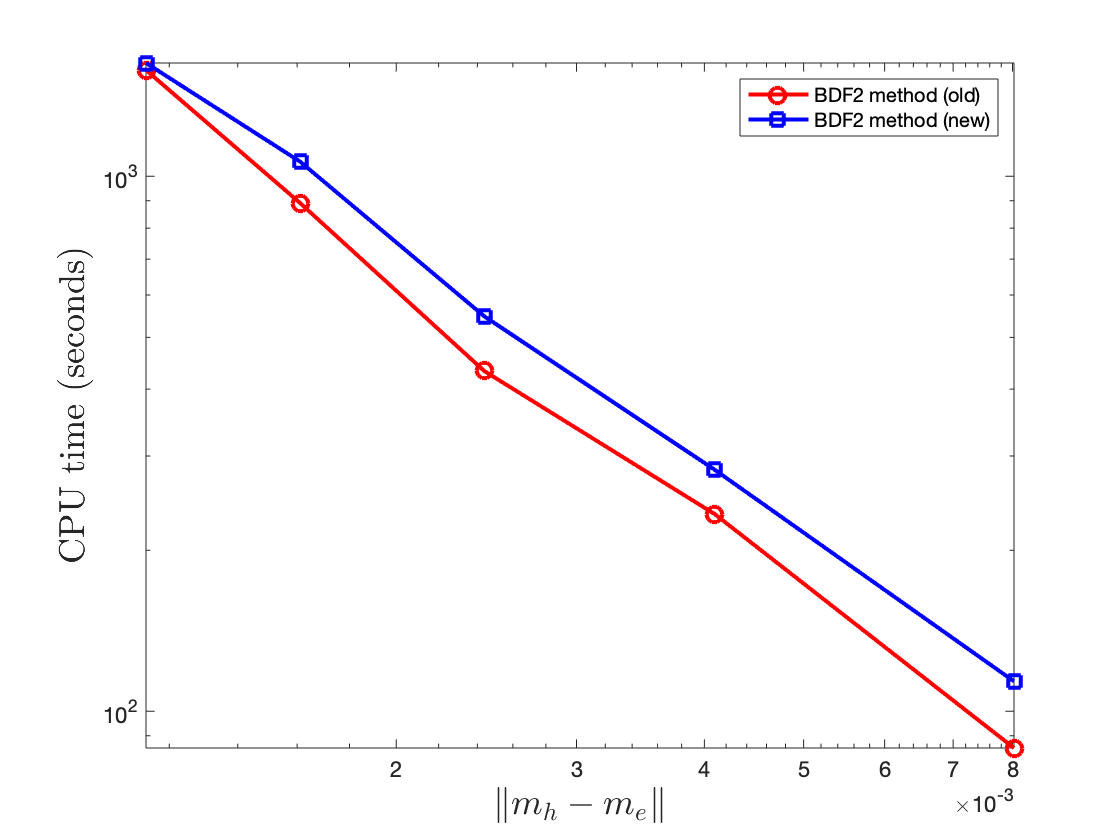}}
		\hspace{0.1in}
		\subfloat[Varying $k$ (1D) ]{\label{cputime_ED_BDF1_1D_time}\includegraphics[width=1.5in]{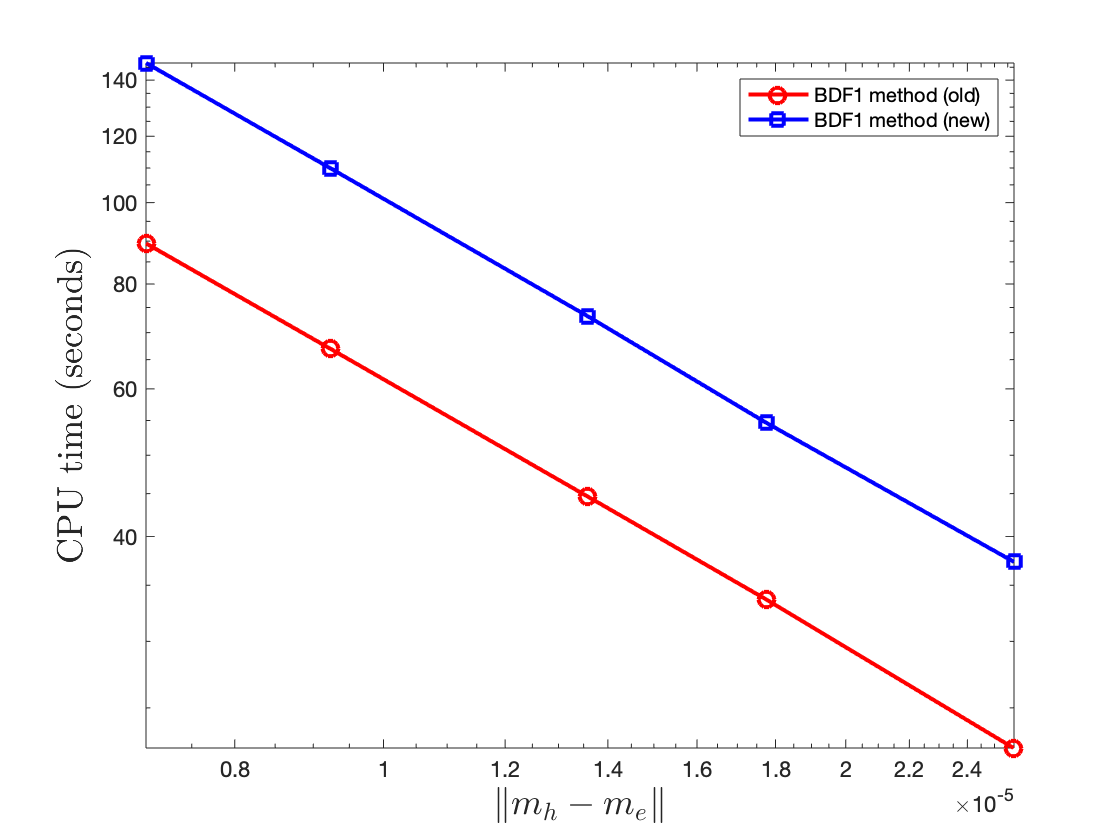}}
		\subfloat[Varying $k$ (3D) ]{\label{cputime_ED_BDF1_3D_time}\includegraphics[width=1.5in]{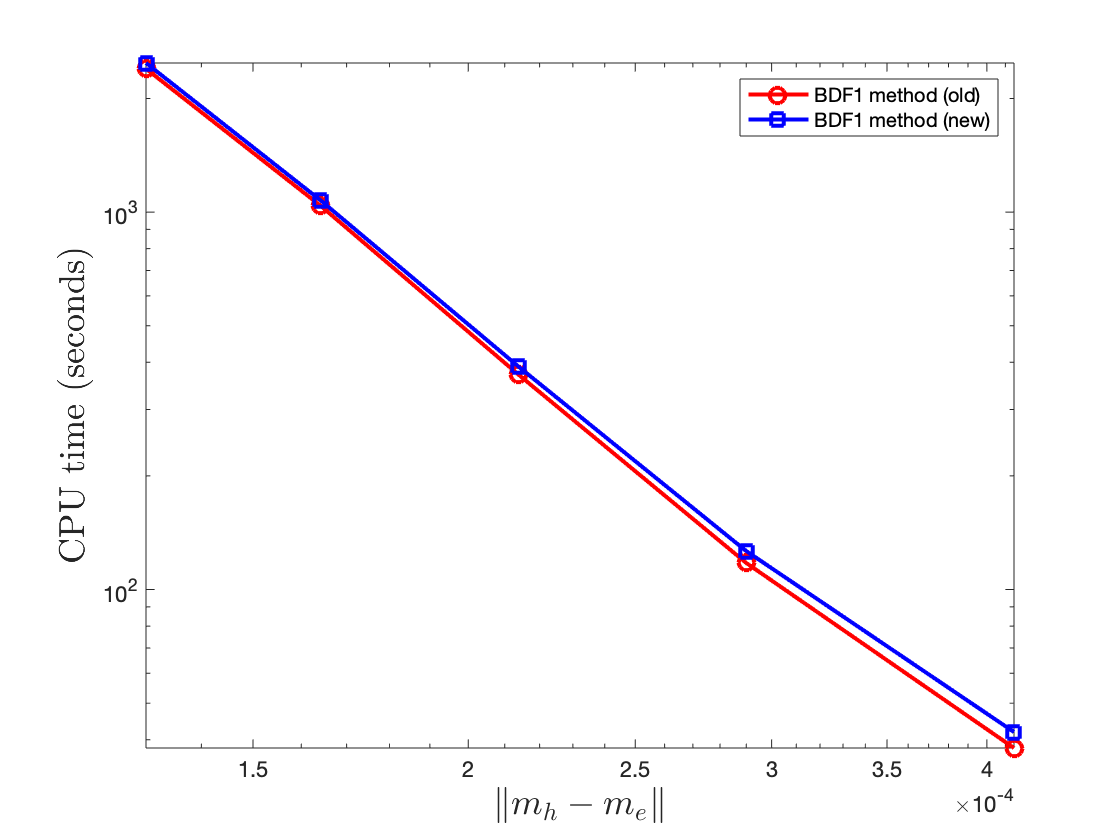}}
		\subfloat[Varying $h$ (1D) ]{\label{cputime_ED_BDF1_1D_space}\includegraphics[width=1.5in]{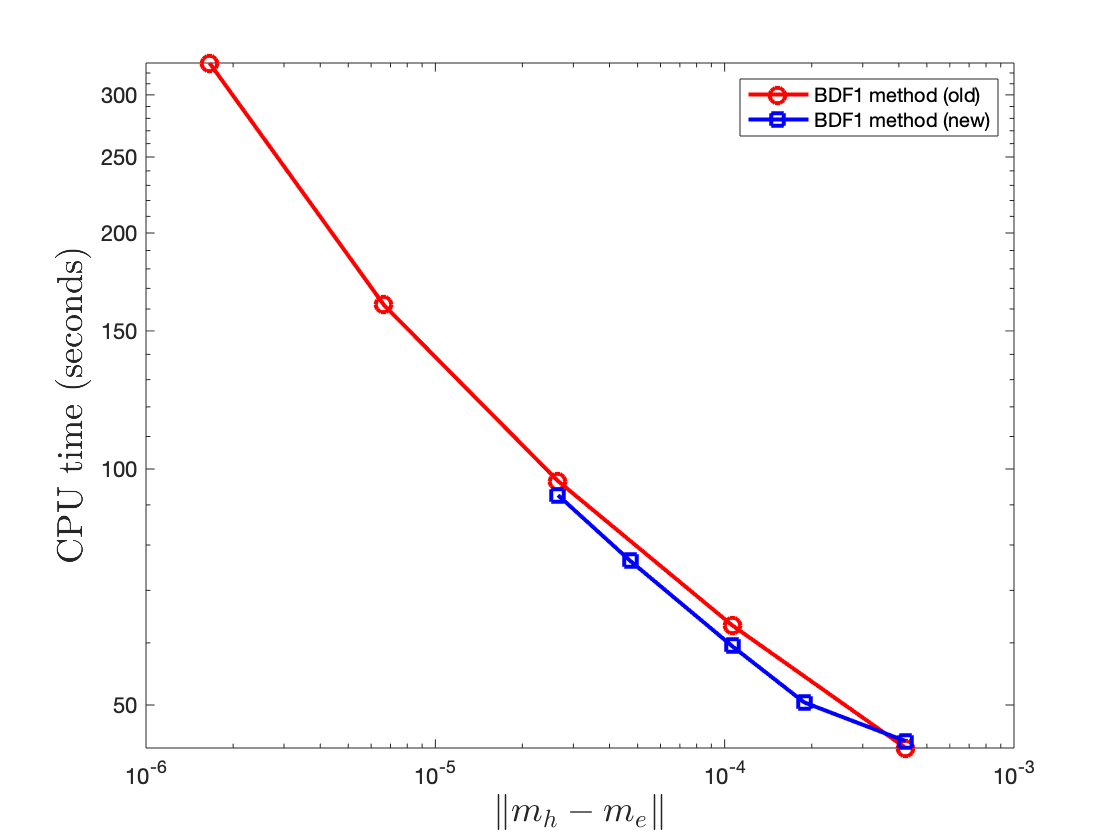}}
		\subfloat[Varying $h$ (3D) ]{\label{cputime_ED_BDF1_3D_space}\includegraphics[width=1.5in]{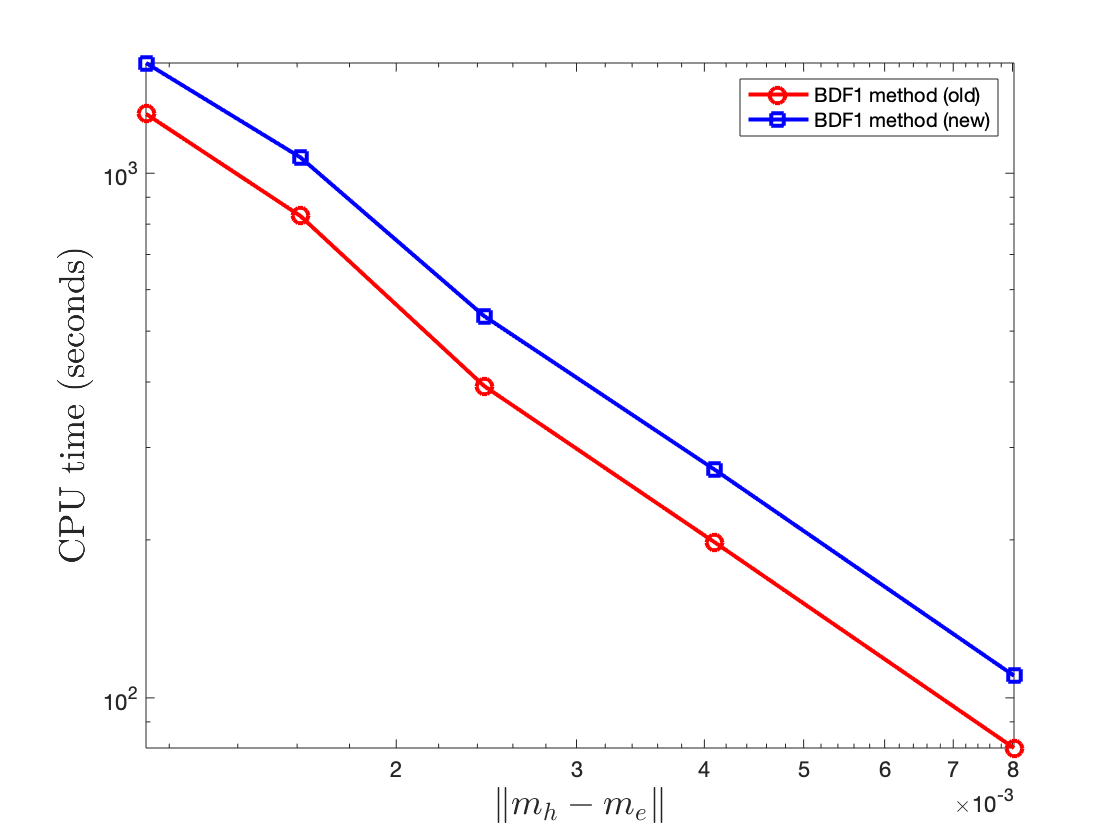}}
	\caption{Comparison of CPU time required to achieve the desired numerical accuracy for the proposed method for \cref{eq-5} and the third order semi-implicit scheme for \cref{eq-model}, along with their BDF2 and BDF1 counterparts. Top row: BDF3 method; Middle row: BDF2 method; Bottom row: BDF1 method. The observation is that the proposed method consumes more time than previous methods to achieve the same level of accuracy, indicating that its linear system is more difficult to solve. Numerical evidence reveals that GMRES solver convergence slows for larger temporal step size $k$ or smaller spatial grid-size $h$, increasing computational challenge. Thus, the proposed method requires a more efficient solver. }\label{figure:3order}
\end{figure}

Additionally, to compare numerical efficiency between the proposed method \cref{proposed} and the revised third-order semi-implicit scheme \cref{scheme-third-order}, we plot CPU time (seconds) versus the $\ell^{\infty}$ norm. Results are presented in \cref{figure:3order}. We observe that the proposed method requires more time than previous methods to achieve the same accuracy level. This indicates that the linear system for the proposed method is more challenging to solve. Numerical evidence shows that GMRES solver convergence slows for larger temporal step size $k$ or smaller spatial grid size $h$, increasing computational difficulty. Thus, the proposed method necessitates a more efficient solver.

\subsection{Stability test}
\begin{figure}[htbp]
	\centering
	\subfloat{\label{BDF3_alpha_0_ang}\includegraphics[width=1.3in]{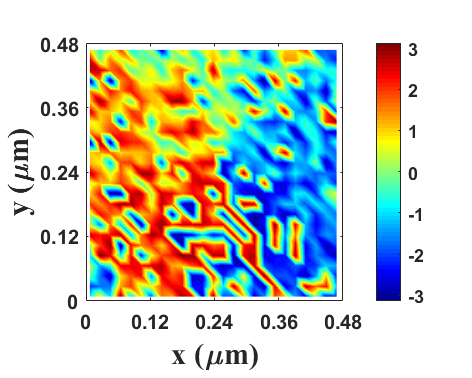}}
	\subfloat{\label{BDF3_alpha_0dot01_ang}\includegraphics[width=1.3in]{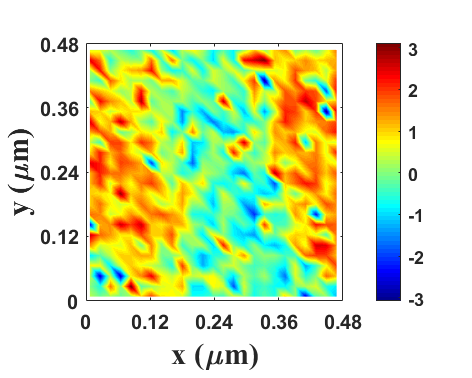}}	\subfloat{\label{BDF3_alpha_0dot1_ang}\includegraphics[width=1.3in]{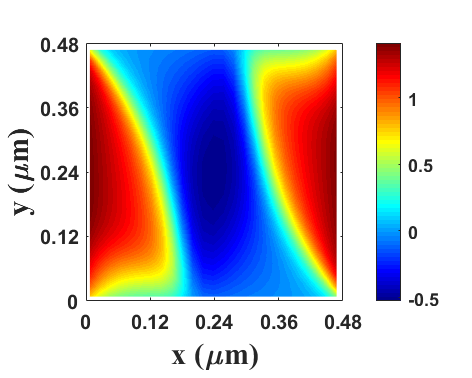}}
	\subfloat{\label{BDF3_alpha_1_ang}\includegraphics[width=1.3in]{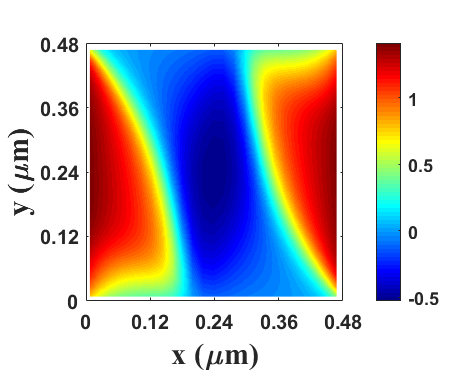}}
	\hspace{0.1in}
	\subfloat{\label{BDF3_alpha_5_ang}\includegraphics[width=1.3in]{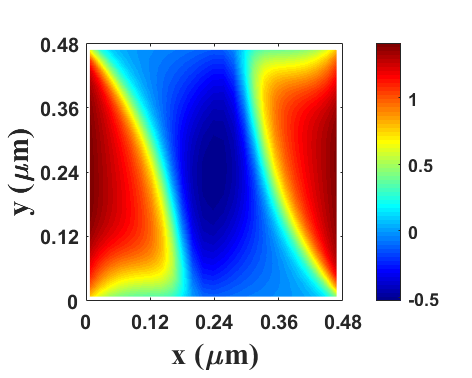}}
	\subfloat{\label{BDF3_alpha_10_ang}\includegraphics[width=1.3in]{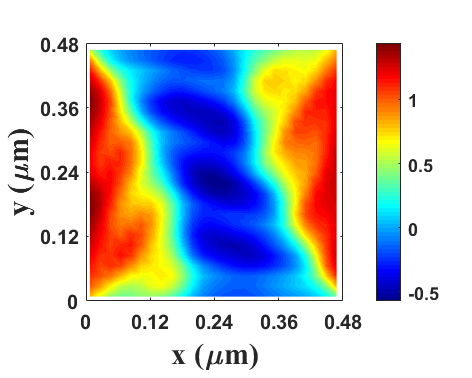}}
	\subfloat{\label{BDF3_alpha_40_ang}\includegraphics[width=1.3in]{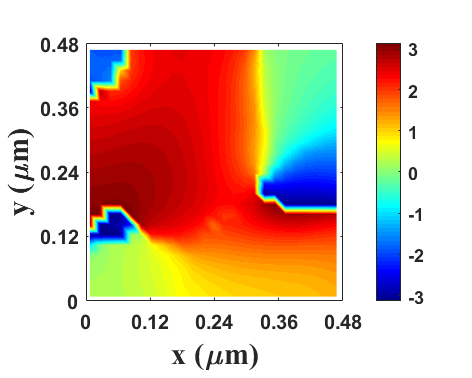}}
	\subfloat{\label{BDF3_alpha_100_ang}\includegraphics[width=1.3in]{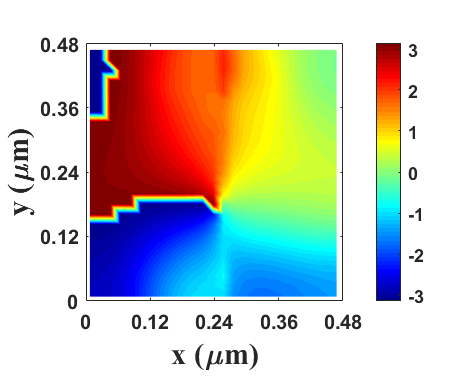}}
		\hspace{0.1in}
	\subfloat{\label{BDF2_alpha_0_ang}\includegraphics[width=1.3in]{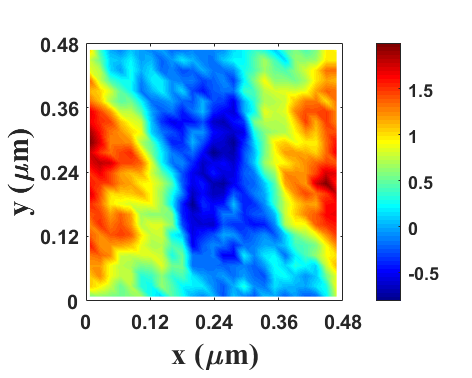}}
	\subfloat{\label{BDF2_alpha_0dot01_ang}\includegraphics[width=1.3in]{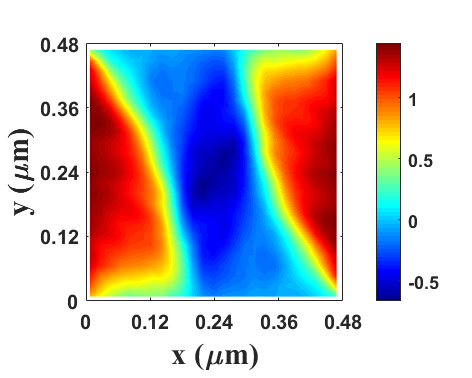}}
	\subfloat{\label{BDF2_alpha_0dot1_ang}\includegraphics[width=1.3in]{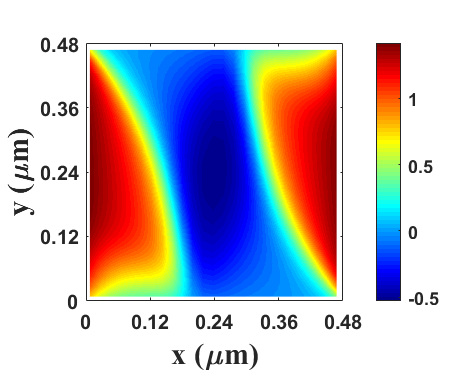}}
	\subfloat{\label{BDF2_alpha_1_ang}\includegraphics[width=1.3in]{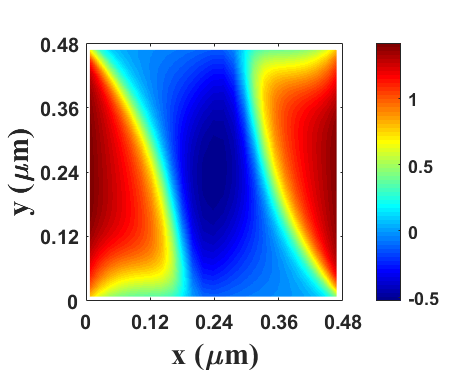}}
	\hspace{0.1in}
	\subfloat{\label{BDF2_alpha_5_ang}\includegraphics[width=1.3in]{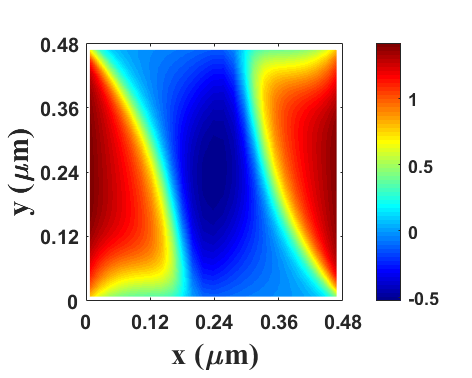}}
	\subfloat{\label{BDF2_alpha_10_ang}\includegraphics[width=1.3in]{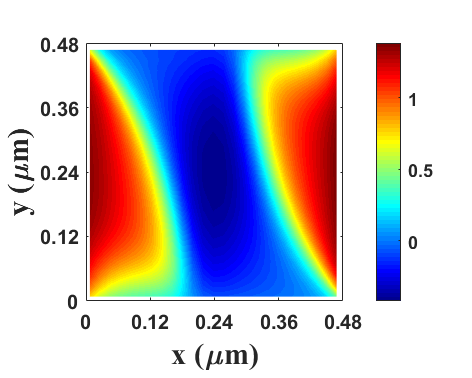}}
	\subfloat{\label{BDF2_alpha_40_ang}\includegraphics[width=1.3in]{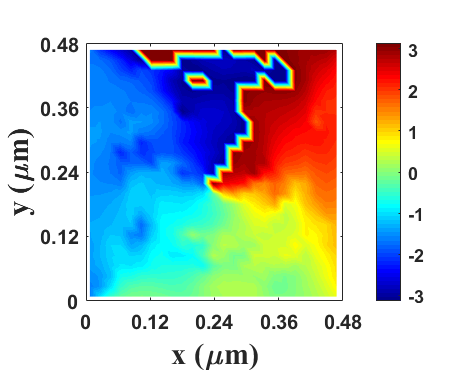}}
	\subfloat{\label{BDF2_alpha_100_ang}\includegraphics[width=1.3in]{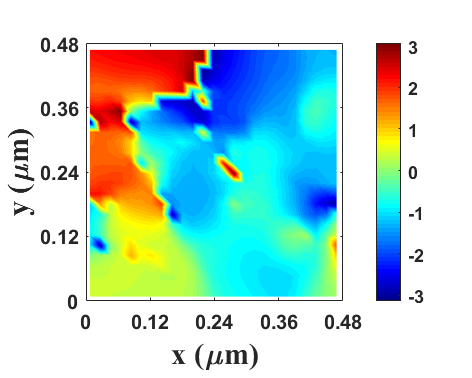}}
		\hspace{0.1in}
	\subfloat{\label{BDF1_alpha_0_ang}\includegraphics[width=1.3in]{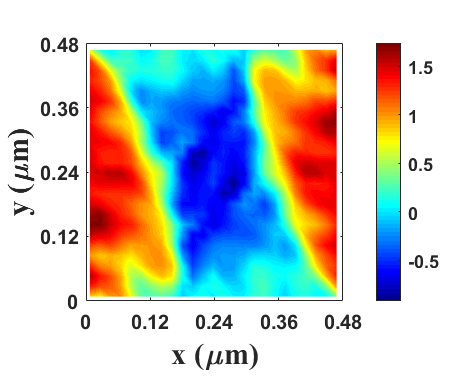}}
	\subfloat{\label{BDF1_alpha_0dot01_ang}\includegraphics[width=1.3in]{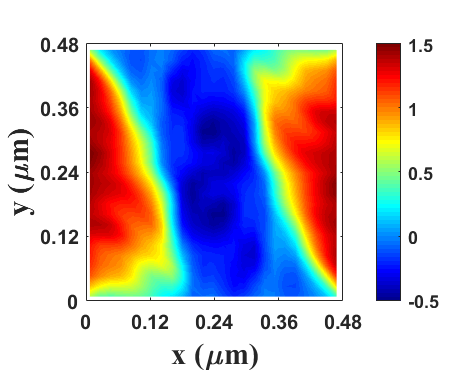}}
	\subfloat{\label{BDF1_alpha_0dot1_ang}\includegraphics[width=1.3in]{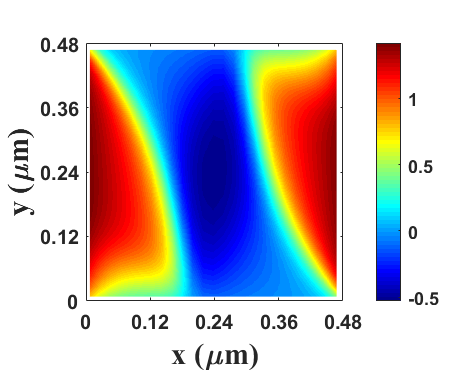}}
	\subfloat{\label{BDF1_alpha_1_ang}\includegraphics[width=1.3in]{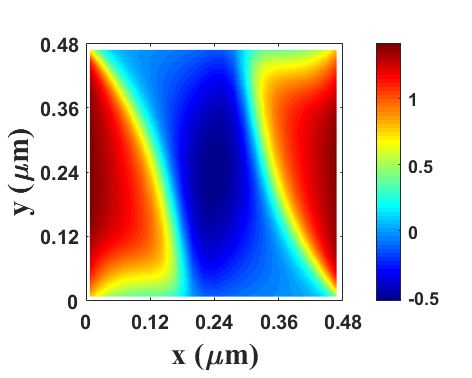}}
	\hspace{0.1in}
	\subfloat{\label{BDF1_alpha_5_ang}\includegraphics[width=1.3in]{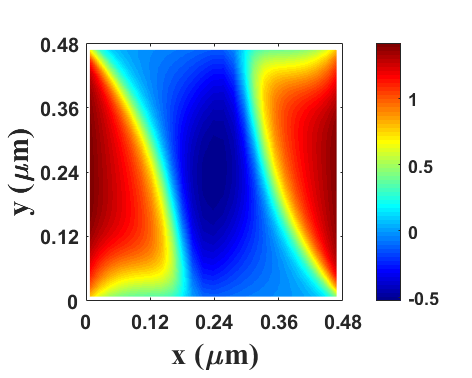}}
	\subfloat{\label{BDF1_alpha_10_ang}\includegraphics[width=1.3in]{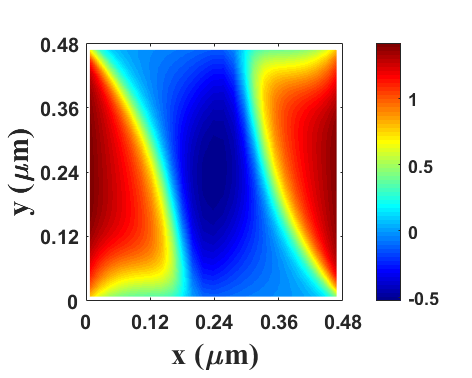}}
	\subfloat{\label{BDF1_alpha_40_ang}\includegraphics[width=1.3in]{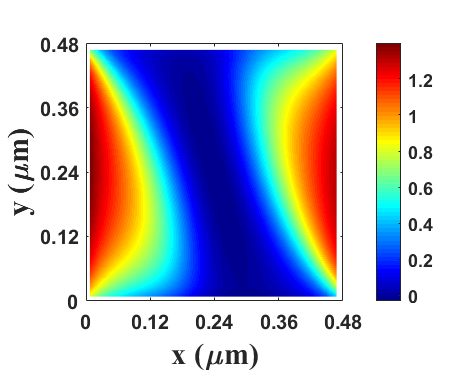}}
	\subfloat{\label{BDF1_alpha_100_ang}\includegraphics[width=1.3in]{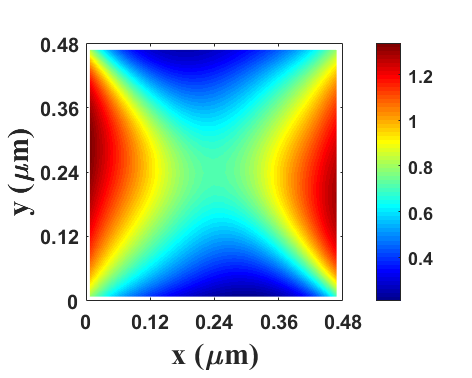}}
	\caption{Stable structures in the absence of magnetic field at $2\,$ns.  Color denotes the angle between the first two components of the magnetization vector. Top two rows: Proposed method; Middle two rows: BDF2; Bottom two rows: BDF1. From left to right: $\alpha=0,0.01,0.1,1,5,10,40,100$. $k=1\;ps$. }\label{BDF3_BDF2_BDF1_alpha}
\end{figure}

\begin{figure}[htbp]
	\centering
	\subfloat{\label{BDF3_alpha_0_ang_v1}\includegraphics[width=1.3in]{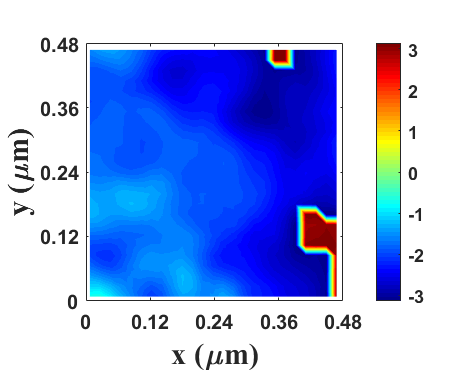}}
	\subfloat{\label{BDF3_alpha_0dot01_ang_v1}\includegraphics[width=1.3in]{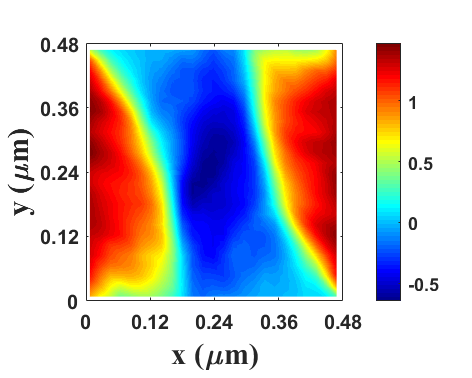}}	\subfloat{\label{BDF3_alpha_0dot1_ang_v1}\includegraphics[width=1.3in]{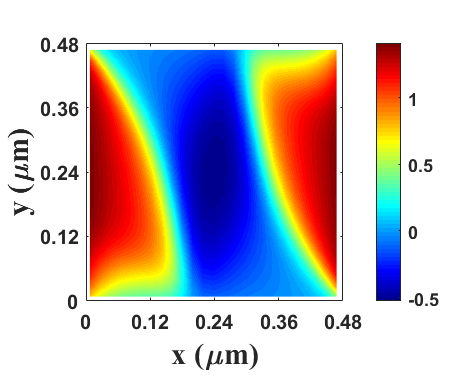}}
	\subfloat{\label{BDF3_alpha_1_ang_v1}\includegraphics[width=1.3in]{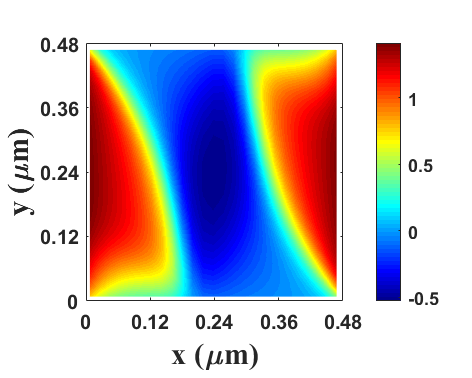}}
	\hspace{0.1in}
	\subfloat{\label{BDF3_alpha_5_ang_v1}\includegraphics[width=1.3in]{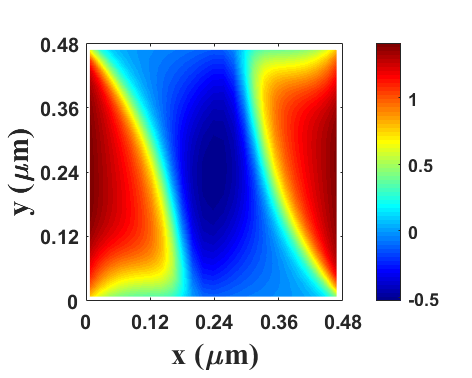}}
	\subfloat{\label{BDF3_alpha_10_ang_v1}\includegraphics[width=1.3in]{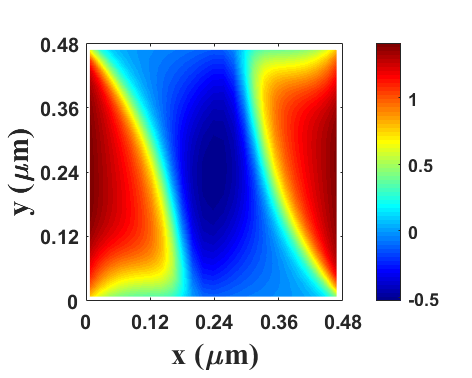}}
	\subfloat{\label{BDF3_alpha_40_ang_v1}\includegraphics[width=1.3in]{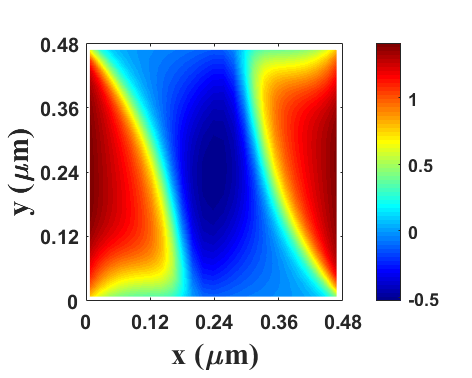}}
	\subfloat{\label{BDF3_alpha_100_ang_v1}\includegraphics[width=1.3in]{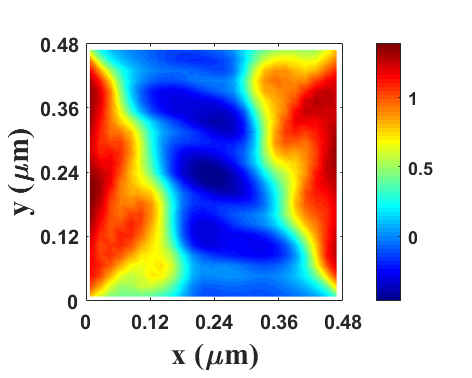}}
		\hspace{0.1in}
	\subfloat{\label{BDF2_alpha_0_ang_v1}\includegraphics[width=1.3in]{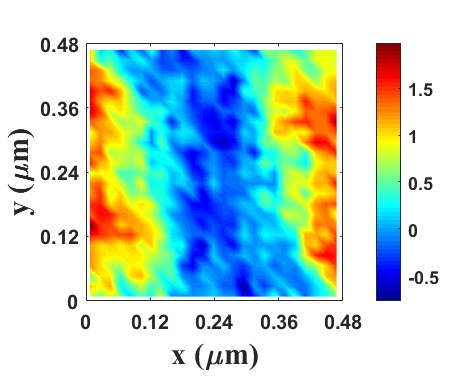}}
	\subfloat{\label{BDF2_alpha_0dot01_ang_v1}\includegraphics[width=1.3in]{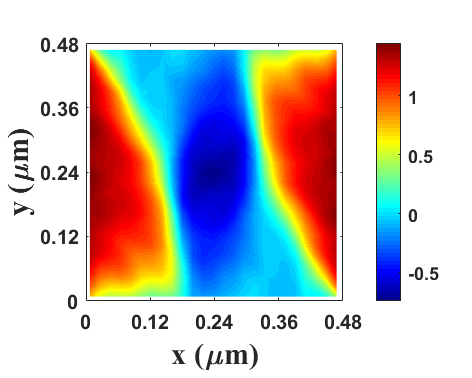}}
	\subfloat{\label{BDF2_alpha_0dot1_ang_v1}\includegraphics[width=1.3in]{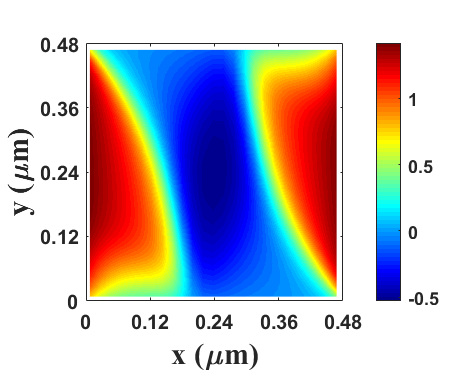}}
	\subfloat{\label{BDF2_alpha_1_ang_v1}\includegraphics[width=1.3in]{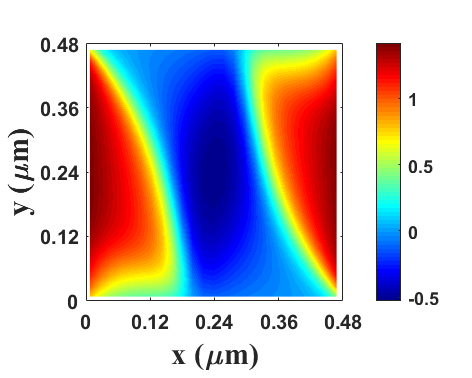}}
	\hspace{0.1in}
	\subfloat{\label{BDF2_alpha_5_ang_v1}\includegraphics[width=1.3in]{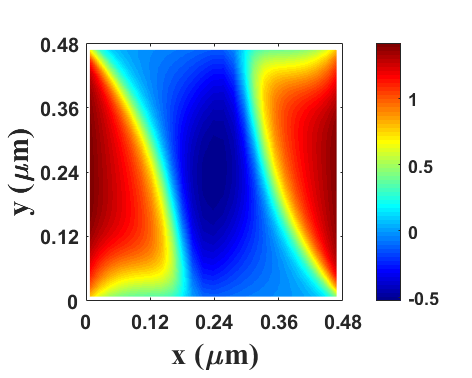}}
	\subfloat{\label{BDF2_alpha_10_ang_v1}\includegraphics[width=1.3in]{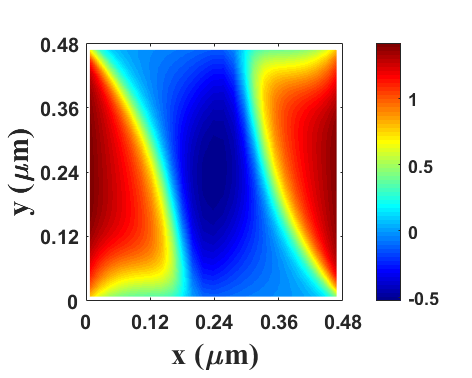}}
	\subfloat{\label{BDF2_alpha_40_ang_v1}\includegraphics[width=1.3in]{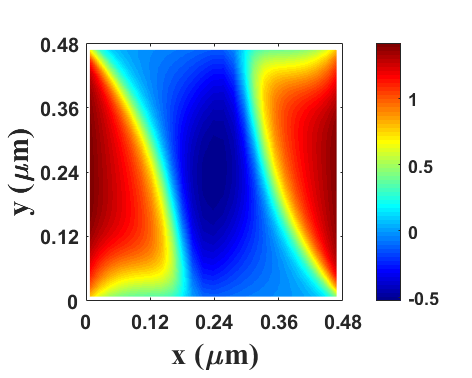}}
	\subfloat{\label{BDF2_alpha_100_ang_v1}\includegraphics[width=1.3in]{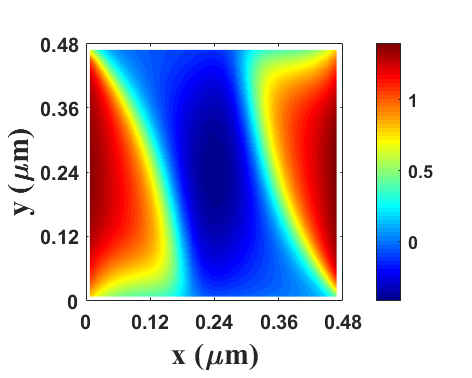}}
	\hspace{0.1in}
	\subfloat{\label{BDF1_alpha_0_ang_v1}\includegraphics[width=1.3in]{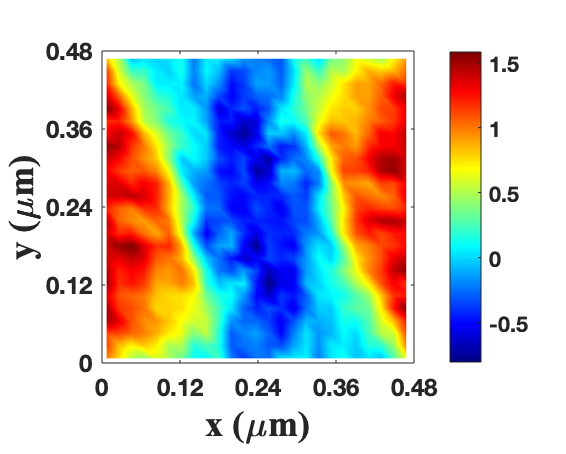}}
	\subfloat{\label{BDF1_alpha_0dot01_ang_v1}\includegraphics[width=1.3in]{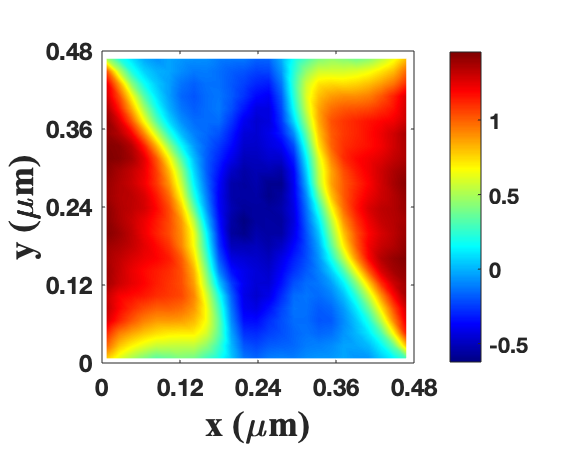}}
	\subfloat{\label{BDF1_alpha_0dot1_ang_v1}\includegraphics[width=1.3in]{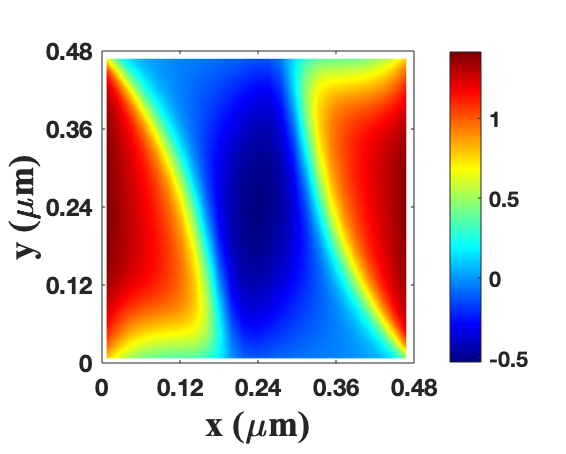}}
	\subfloat{\label{BDF1_alpha_1_ang_v1}\includegraphics[width=1.3in]{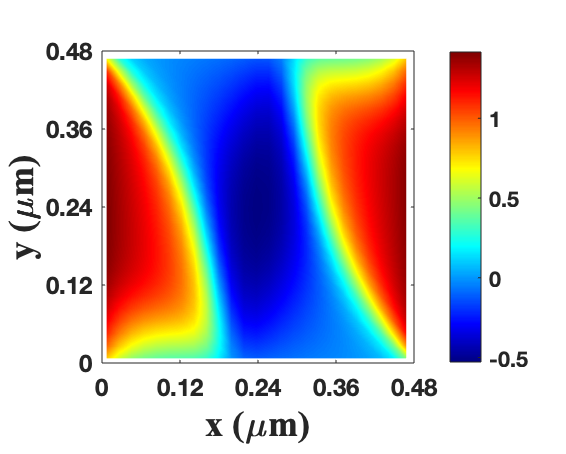}}
		\hspace{0.1in}
	\subfloat{\label{BDF1_alpha_5_ang_v1}\includegraphics[width=1.3in]{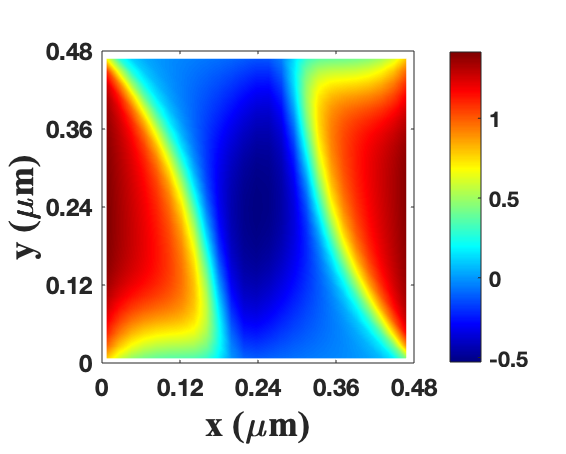}}
	\subfloat{\label{BDF1_alpha_10_ang_v1}\includegraphics[width=1.3in]{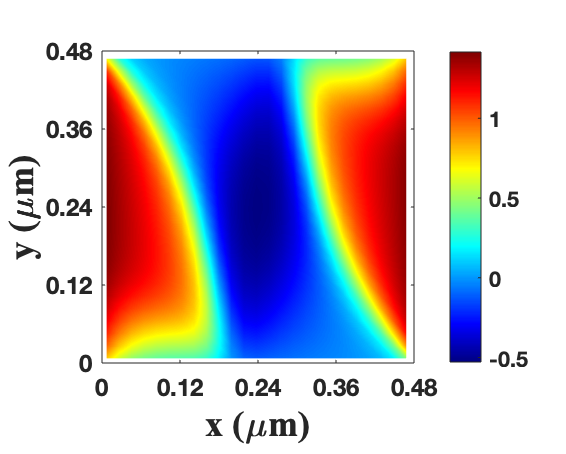}}
	\subfloat{\label{BDF1_alpha_40_ang_v1}\includegraphics[width=1.3in]{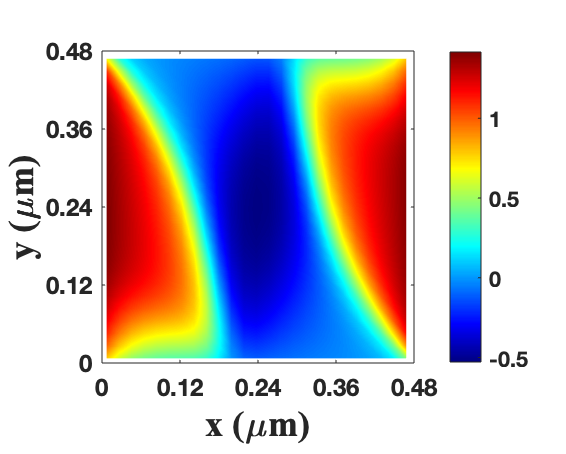}}
	\subfloat{\label{BDF1_alpha_100_ang_v1}\includegraphics[width=1.3in]{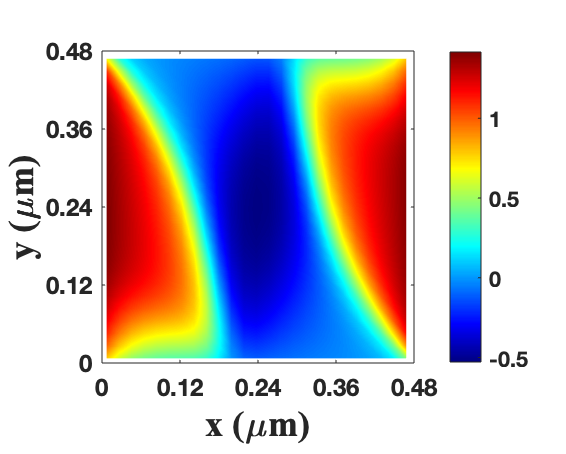}}
	\caption{Stable structures in the absence of magnetic field at $2\,$ns.  Color denotes the angle between the first two components of the magnetization vector. Top two rows: Proposed method; Middle two rows: BDF2; Bottom two rows: BDF1. From left to right: $\alpha=0,0.01,0.1,1,5,10,40,100$. $k=0.1\;ps$. }\label{BDF3_BDF2_BDF1_alpha_v1}
\end{figure}

To evaluate the numerical stability of the three methods in practical micromagnetic simulations with arbitrary damping parameters, a systematic set of numerical experiments was designed. The simulation model uses a ferromagnetic thin film with dimensions $480\times480\times20\,\textrm{nm}^3$, discretized on a $100\times100\times4$ grid—a resolution balancing capture of fine magnetic structures with computational efficiency. Two temporal step sizes, $k=1\,\text{ps}$ and $k=0.1\,\text{ps}$, were used to investigate time discretization influence on stability. The initial magnetization was set uniform along the $x$-direction, and the external magnetic field was zero to eliminate external driving effects, thereby isolating the impact of damping parameters on system stability. A broad range of damping parameter values was tested to cover weak, moderate, and extremely strong dissipation scenarios: $\alpha=0,0.01,0.1,1,5,10,40,100$. The resulting magnetization profiles, which confirm stable evolution under all tested conditions, are presented in \cref{BDF3_BDF2_BDF1_alpha} and \cref{BDF3_BDF2_BDF1_alpha_v1}. Key observations from these comprehensive simulations are as follows.
\begin{itemize}
	\item The proposed method becomes unstable for very large damping parameters, whereas BDF2 yields reasonable configurations and BDF1 produces different stable results;
	\item All three methods remain stable for moderately large $\alpha$;
	\item The proposed method is the only one that exhibits instability for small$\alpha$.
	\item Stability is generally improved with smaller time steps $k$.
	\item The results by proposed method with $\alpha=10$ and BDF2 method with $\alpha=40$ are less stable than the third order scheme \cref{scheme-third-order} in our previous work \cite{xie2025thirdorder}. However, the results of BDF1 method for \cref{eq-5} are much more stable than that of BDF1 method for \cref{eq-model} in \cite{xie2025thirdorder}.
\end{itemize}

A preliminary theoretical analysis of the proposed numerical method provides insight into its convergence performance. This theoretical threshold offers a mathematical basis for understanding the method's convergence behavior under moderate-to-strong damping conditions. Complementing the theoretical findings, an extensive set of numerical experiments has been conducted to validate the method's practical stability characteristics. These experiments, covering a wide range of damping parameters and simulation configurations, consistently suggest that a more relaxed condition \(\alpha>0.1\) is sufficient to ensure robust numerical stability in practical micromagnetic computations.

\begin{figure}[htbp]
	\centering
	\subfloat[Proposed]{\label{energy_BDF3}\includegraphics[width=1.8in]{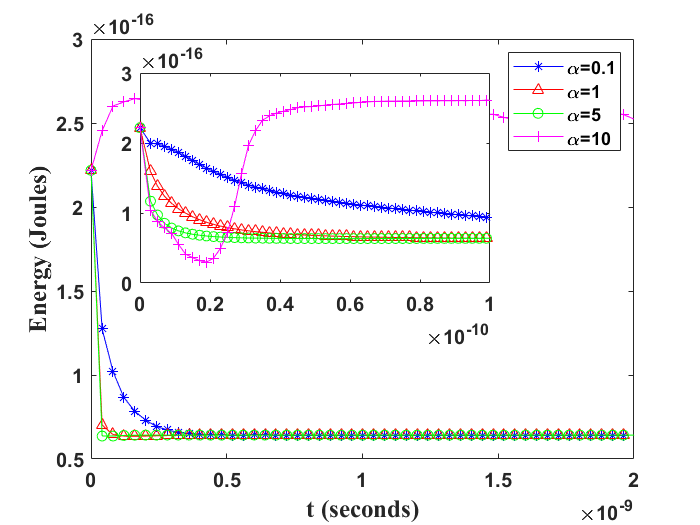}}
	\subfloat[BDF1]{\label{energy_BDF1}\includegraphics[width=1.8in]{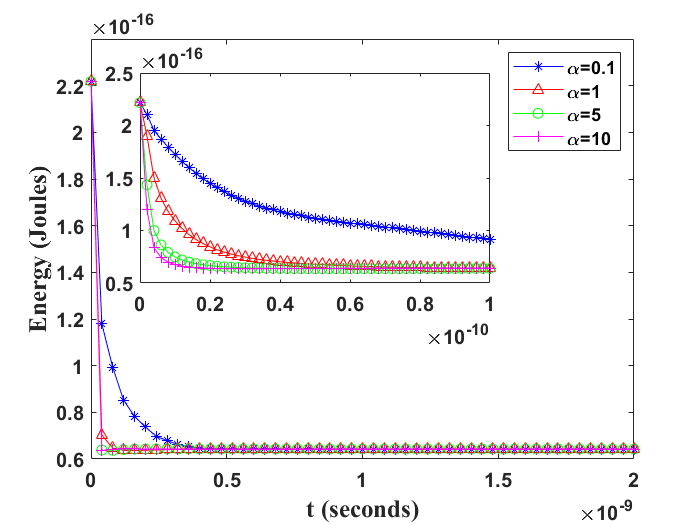}}
	\subfloat[BDF2]{\label{energy_BDF2}\includegraphics[width=1.8in]{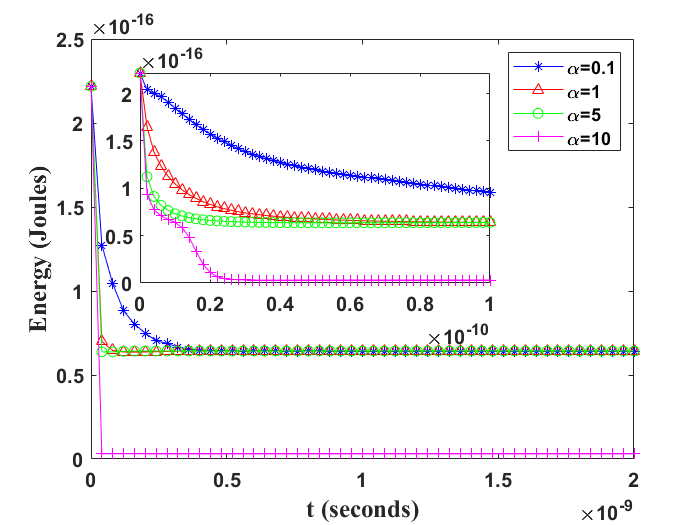}}
	\caption{Energy evolution curves of three numerical methods, with different damping constants, $\alpha=0.1,1,5,10$, up to $t=2\,$ns  in the absence of external magnetic field. Left: Proposed numerical method; Middle: BDF1; Right: BDF2. One common feature is that the energy dissipation rate is faster for larger $\alpha$, which is physically reasonable. The proposed method is unstable with $\alpha=10$.}\label{energy_decay}
\end{figure}

Using the same simulation setup detailed previously, we performed a comparative study of the energy dissipation characteristics of the three numerical methods: the proposed scheme, BDF2, and BDF1. The magnetic system reaches a stable state by \(t=2\,\text{ns}\) across all tested configurations, and the total magnetic energy throughout the simulation is calculated using the energy formulation in \eqref{LL-Energy}.
Energy evolution curves for the three methods, each tested under four damping parameter values (\(\alpha=0.1,1,5,10\)), are presented in \cref{energy_decay}. A prominent and consistent feature observed across all schemes is that the energy dissipation rate increases with the Gilbert damping parameter \(\alpha\), except for the unstable case of the proposed method under $\alpha=10$. Specifically, simulations with larger \(\alpha\) values exhibit a more rapid decline in total magnetic energy until stabilization, consistent with the physical expectation that higher damping facilitates faster energy dissipation. Consequently, the energy evolution trends captured by BDF1, BDF2, and the proposed method all show excellent agreement with theoretical analysis of the LLG equation, confirming the physical consistency and reliability of the three numerical schemes.

\begin{figure}[htbp]
	\centering
	\subfloat[$\alpha=0.1$]{\label{alpha_0dot1}\includegraphics[width=2.5in]{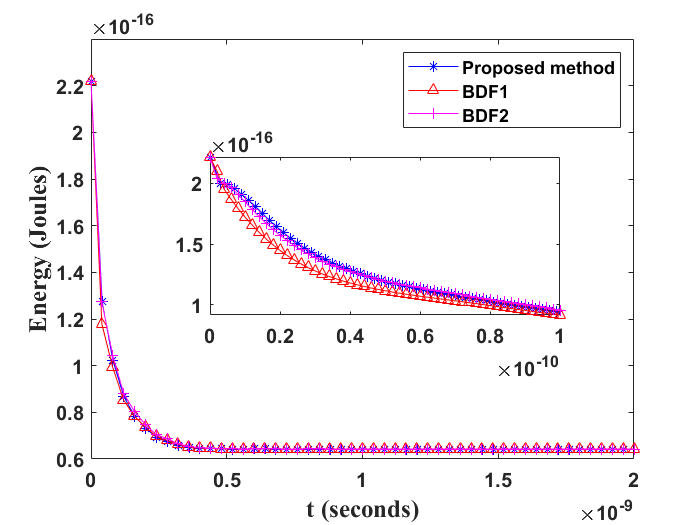}}
	\subfloat[$\alpha=1$]{\label{alpha_1}\includegraphics[width=2.5in]{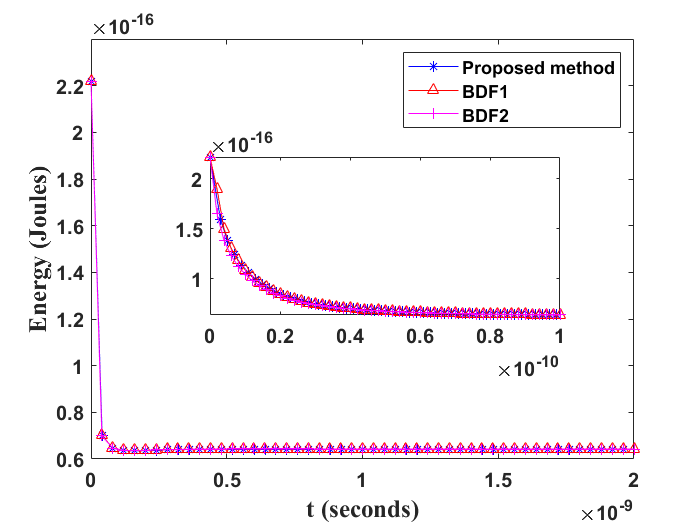}}
	\hspace{0.1in} 
	\subfloat[$\alpha=5$]{\label{alpha_5}\includegraphics[width=2.5in]{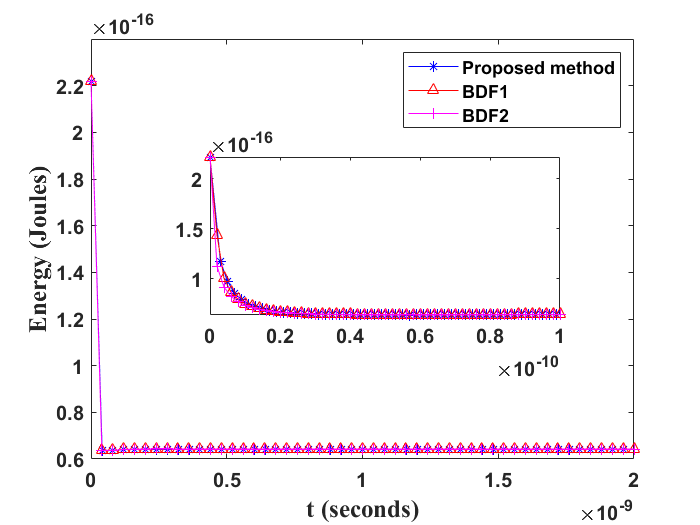}}
	\subfloat[$\alpha=10$]{\label{alpha_10}\includegraphics[width=2.5in]{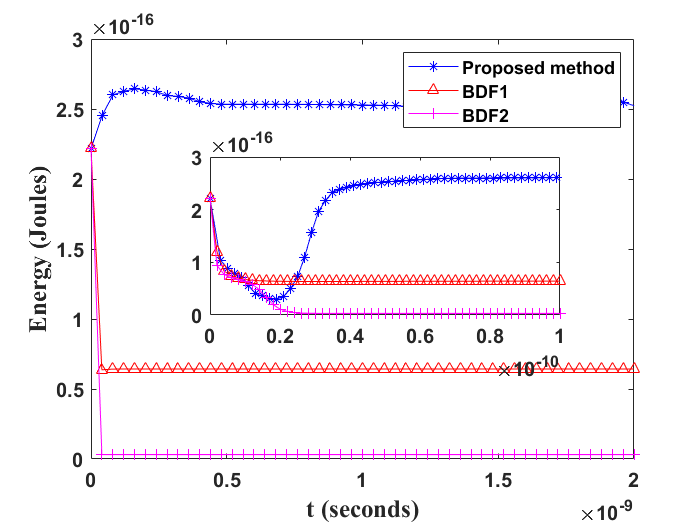}}
	\caption{Energy evolution curves in terms of time, for the numerical results created by three numerical methods up to $t=2\,$ns in the absence of external magnetic field for (a) $\alpha=0.1$, (b) $\alpha=1$, (c) $\alpha=5$, and (d) $\alpha=10$. The energy dissipation of the proposed method is consistent with that of the BDF2 and the BDF1 under $\alpha=0.1,1,5$, but inconsistent for $\alpha=10$.}\label{energy_decay_alpha}
\end{figure}

\begin{figure}[htbp]
	\centering
	\subfloat{\label{alpha_0dot1_ED_BDF3}\includegraphics[width=1.5in]{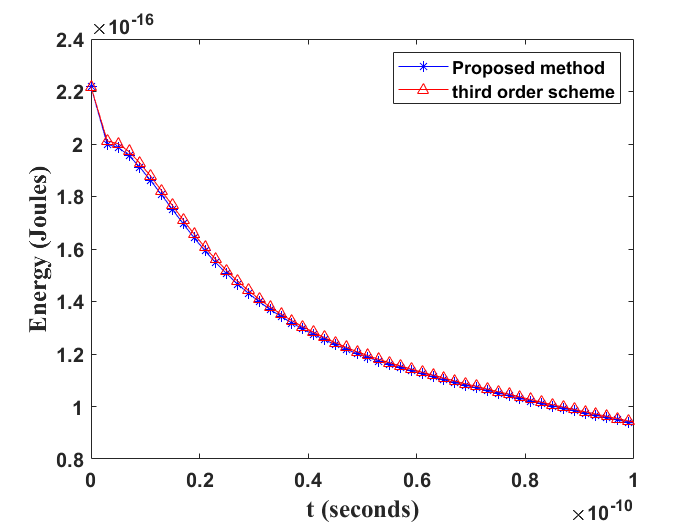}}
		\subfloat{\label{alpha_1_ED_BDF3}\includegraphics[width=1.5in]{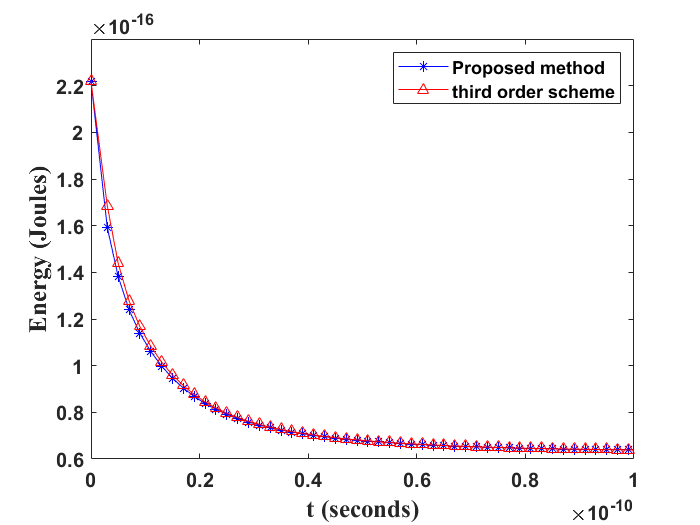}}
		\subfloat{\label{alpha_5_ED_BDF3}\includegraphics[width=1.5in]{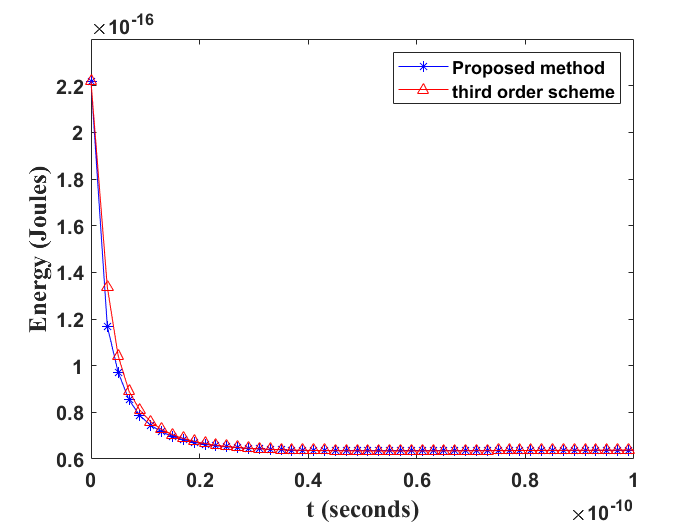}}
		\subfloat{\label{alpha_10_ED_BDF3}\includegraphics[width=1.5in]{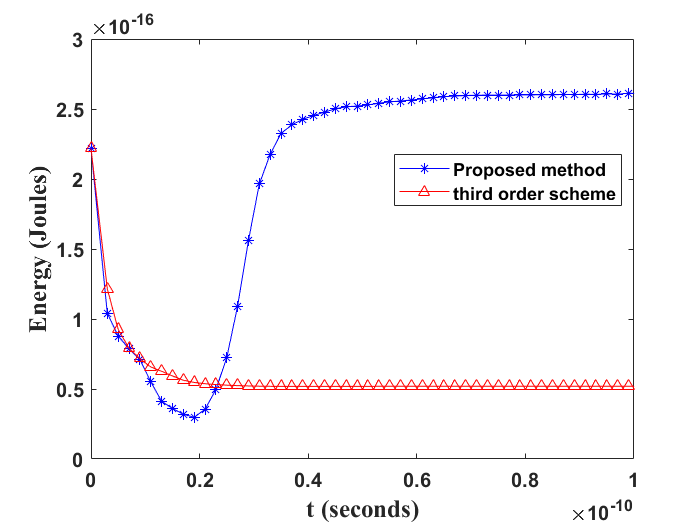}}
		\hspace{0.1in}
	\subfloat{\label{alpha_0dot1_ED_BDF2}\includegraphics[width=1.5in]{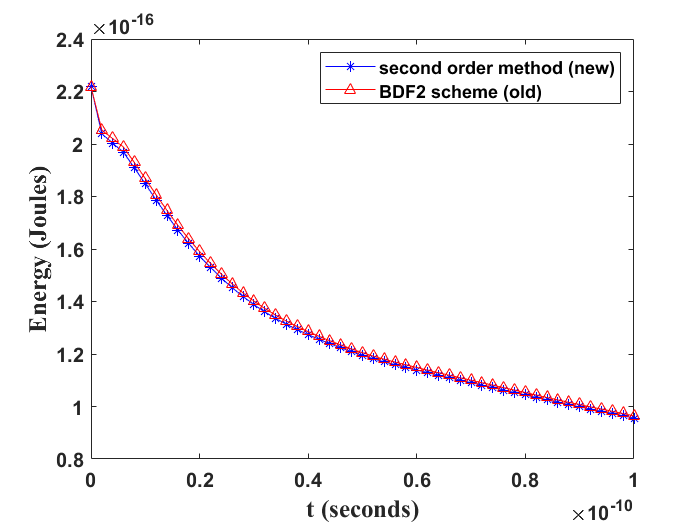}}
	\subfloat{\label{alpha_1_ED_BDF2}\includegraphics[width=1.5in]{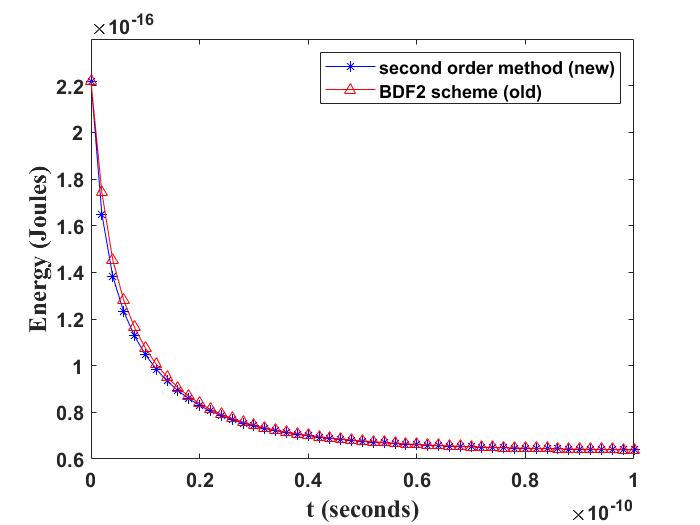}}
	\subfloat{\label{alpha_5_ED_BDF2}\includegraphics[width=1.5in]{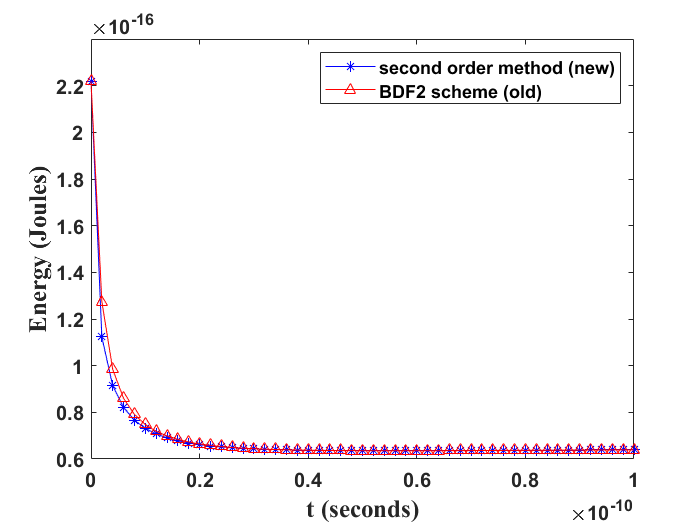}}
	\subfloat{\label{alpha_10_ED_BDF2}\includegraphics[width=1.5in]{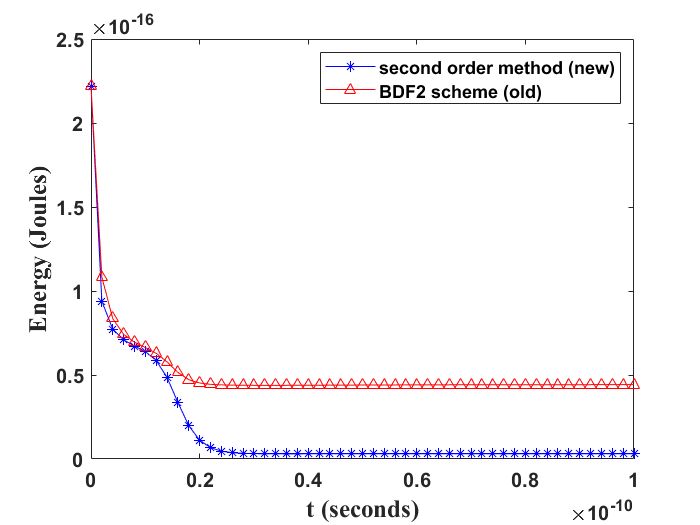}}
	\hspace{0.1in}
	\subfloat{\label{alpha_0dot1_ED_BDF1}\includegraphics[width=1.5in]{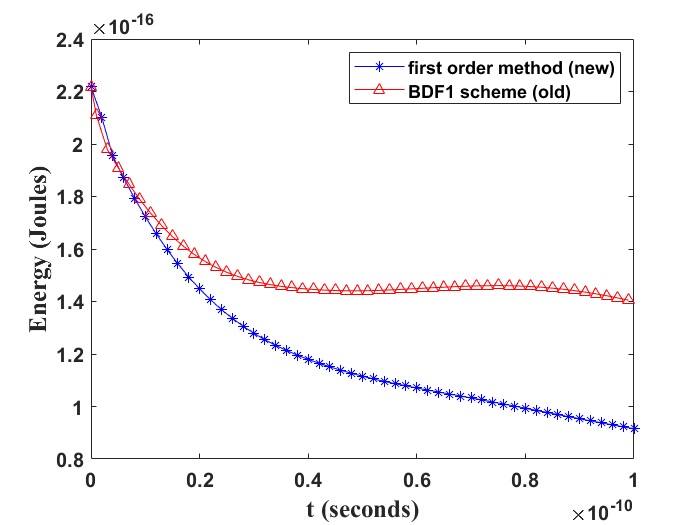}}
	\subfloat{\label{alpha_1_ED_BDF1}\includegraphics[width=1.5in]{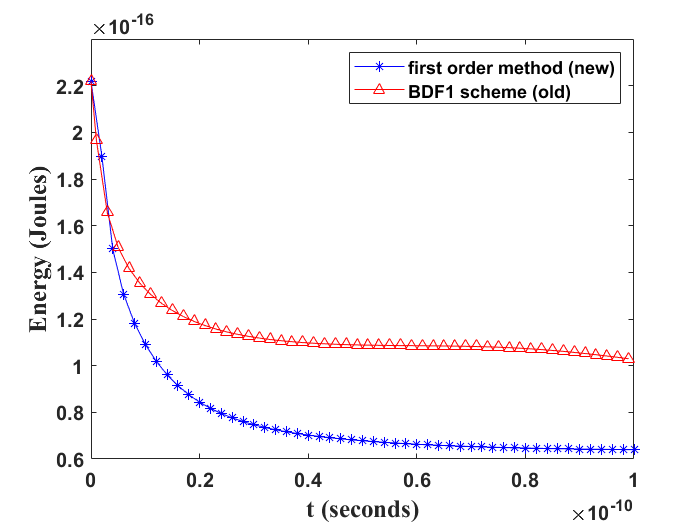}}
	\subfloat{\label{alpha_5_ED_BDF1}\includegraphics[width=1.5in]{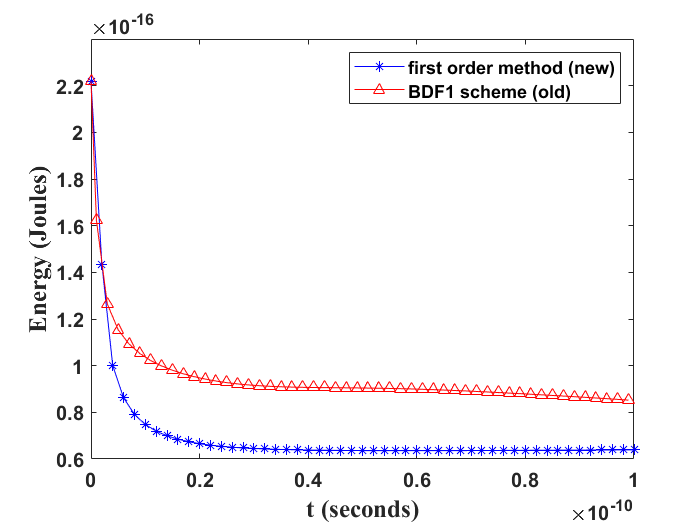}}
	\subfloat{\label{alpha_10_ED_BDF1}\includegraphics[width=1.5in]{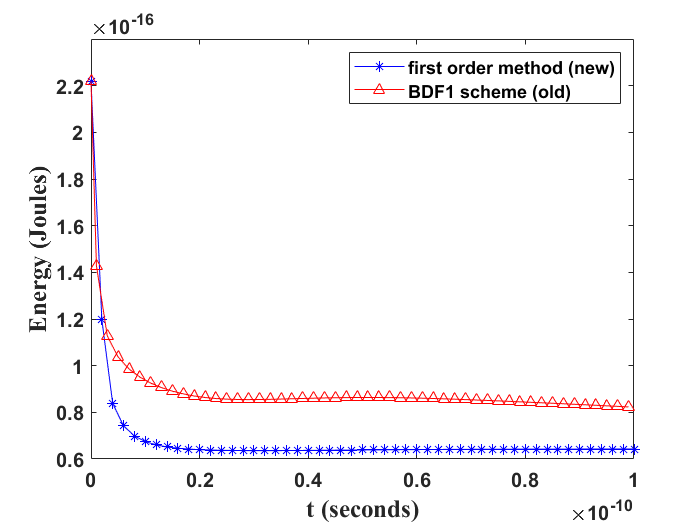}}
	\caption{Comparison for the proposed method \cref{proposed} and a third order semi-implicit scheme \cref{scheme-third-order} and their BDF2 and BDF1 corresponding version. Top row: BDF3 method; Middle row: BDF2 method; Bottom row: BDF1 method. From left to right panel: $\alpha=0.1, 1,5, 10$. The observation is that the proposed methods for \cref{eq-5} achieve much lower energy level than the methods for \cref{eq-model}. The proposed third order method is unstable with $\alpha=10$. }\label{energy_decay_alpha_ED}
\end{figure}

To further elaborate on the energy dissipation characteristics, we used the same sequence of Gilbert damping parameters (\(\alpha\)) as in the preceding analysis. Corresponding energy evolution curves, tracking total magnetic energy over time up to \(T=2\,\text{ns}\) (when the system stabilizes), are presented in  \cref{energy_decay_alpha}.
A detailed comparison reveals several notable patterns. For \(\alpha=0.1,1,5\), the energy dissipation pattern of the proposed method is consistent with that of BDF2 and BDF1, indicating agreement under weak-to-moderate damping. However, this consistency breaks down at \(\alpha=10\) (strong damping): the energy dissipation pattern of the proposed method deviates from both BDF1 and BDF2, yielding an unstable result. Among the three methods, BDF2 exhibits the slowest energy dissipation rate with \(\alpha=10\). Additionally, a comparison between the proposed method \cref{proposed} and the third order semi-implicit scheme \cref{scheme-third-order}, along with their BDF2 and BDF1 counterparts, is shown in \cref{energy_decay_alpha_ED}. One can observe that the proposed methods for \cref{eq-5} achieve much lower energy levels than the methods for \cref{eq-model}. The proposed third order method is unstable with $\alpha=10$. In summary, the proposed method is more sensitive to large dissipation coefficients but exhibits higher stability for reasonably selected dissipation coefficient values.

\subsection{Magnetic Domain wall motion}
\begin{figure}[htbp]
	\centering
	\subfloat[Magnetization for initial state]{\label{NeelWall_initial_mag}\includegraphics[width=2.8in]{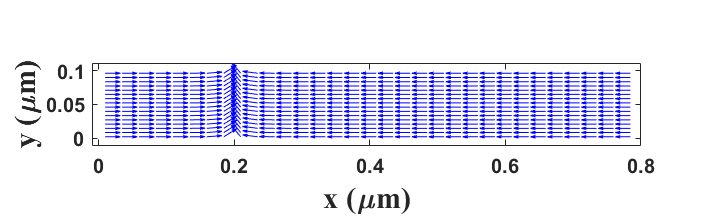}}
	\subfloat[Magnetization with $\alpha=0.1$  at $1.6\,$ns]{\label{NeelWall_alpha_0dot1_mag}\includegraphics[width=2.8in]{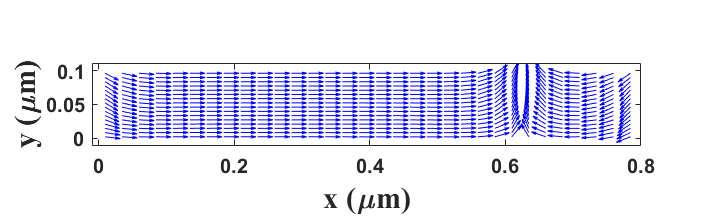}}
	\hspace{0.1in}
	\subfloat[Magnetization with $\alpha=0.4$ at $1.6\,$ns]{\label{NeelWall_alpha_0dot4_mag}\includegraphics[width=2.8in]{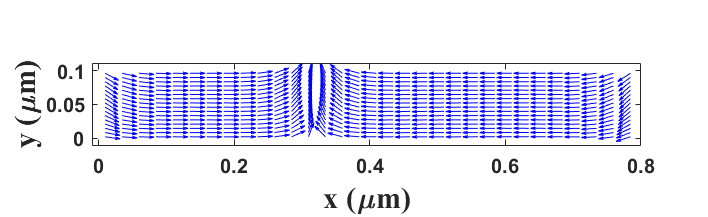}}
	\subfloat[Magnetization with $\alpha=0.8$ at $1.6\,$ns]{\label{NeelWall_alpha_0dot8_mag}\includegraphics[width=2.8in]{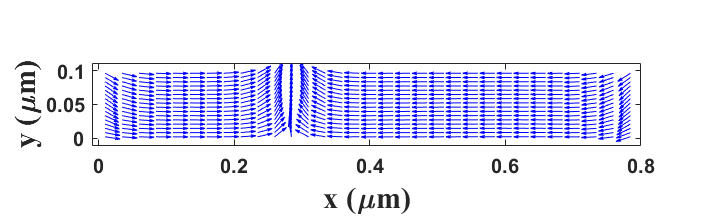}}
	\hspace{0.1in}
	\subfloat[Magnetization with $\alpha=1$ at $1.6\,$ns]{\label{NeelWall_alpha_1_mag}\includegraphics[width=2.8in]{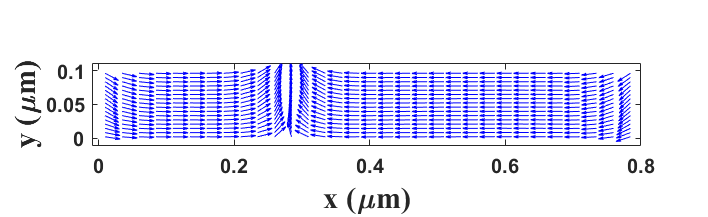}}
	\subfloat[Magnetization with $\alpha=2$ at $1.6\,$ns]{\label{NeelWall_alpha_2_mag}\includegraphics[width=2.8in]{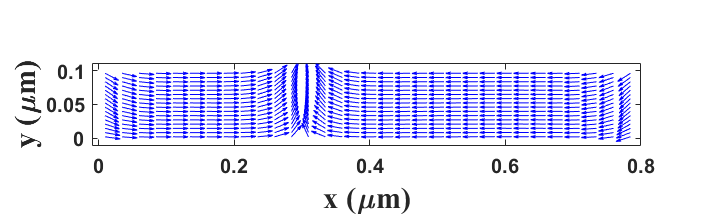}}
	\hspace{0.1in}
	\subfloat[Magnetization with $\alpha=3$ at $1.6\,$ns]{\label{NeelWall_alpha_3_mag}\includegraphics[width=2.8in]{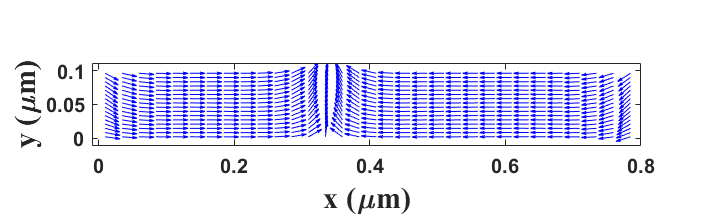}}
	\subfloat[Magnetization with $\alpha=5$ at $1.6\,$ns]{\label{NeelWall_alpha_5_mag}\includegraphics[width=2.8in]{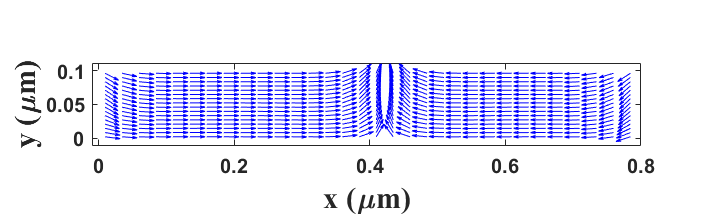}}
	\caption{Magnetization profiles of Ne\'{e}l wall motion in the presence of a magnetic field $\h_e=5\,$mT, with $\alpha = 0.1,0.4,0.8,1,2,3,5$ at $1.6\,$ns for the proposed numerical method. The in-plane arrow denotes the first two components of the magnetization vector. The wall moves slower for larger values of $\alpha$ (when $0<\alpha<1$) and faster for larger values of $\alpha$ (when $\alpha>1$)  and its velocity depends quadratically on $\alpha$. The velocity roughly achieves its minimum value around $\alpha=1$.}\label{NeelWall_alpha_2ns}
\end{figure}

A Neél domain wall was initialized as the initial magnetic state within a ferromagnetic nanostrip of size \(800\times100\times4\,\textrm{nm}^3\). To ensure sufficient spatial resolution for capturing domain wall features while maintaining computational feasibility, the nanostrip was discretized using a structured grid of \(128\times64\times4\) nodes. Following initialization, an external magnetic field of magnitude \(\h_e=5\,\text{mT}\) was applied along the positive \(x\)-direction to induce domain wall motion. Micromagnetic simulations of domain wall dynamics were performed over a time interval of up to \(1.6\,\text{ns}\), The resulting magnetization spatial distributions, which directly reflect the evolution of the Néel wall under different damping conditions, are presented in \cref{NeelWall_alpha_2ns}.
The magnetic domain wall velocity obtained using the proposed method is entirely consistent with that reported in our previous work \cite{xie2025thirdorder}.

\begin{table}[htbp]
	\centering
	{\caption{Consistency of the wall velocity $V$ with respect to the external magnetic field $\h_e$ and the damping parameter $\alpha$ by usage of proposed method \cref{proposed} and our previous third order method \cref{scheme-third-order}. } \label{tab-2-1} }{
	\subfloat[The third-order semi-implicit method \cref{scheme-third-order}]{\label{tab:scheme-third}
		\begin{tabular}{c|c|c|c|c|c|c|c}
			\hline 
			\diagbox{$\h_e(\textrm{mT})$}{$V$ (m/s)}{$\alpha$}&0.1&0.4&0.8 &1 & 2&3&5 \\
			\hline
			5&  227&75  &53  &52  &64 &86&135 \\
			7&313  &106 &76  &74  & 93& 125&192 \\
			9& 385& 139& 98 & 94 & 119& 161&250 \\
			\hline 
		\end{tabular}
	}	
\qquad
\subfloat[The proposed method \cref{proposed}]{\label{tab:proposed}
	\begin{tabular}{c|c|c|c|c|c|c|c}
		\hline 
		\diagbox{$\h_e(\textrm{mT})$}{$V$ (m/s)}{$\alpha$}&0.1&0.4&0.8 &1 & 2&3&5 \\
		\hline
		5&  227&75  &53  &52  &64 &86&135\\
		7&313  &106 &76  &74  & 93& 125&192 \\
		9& 385& 139& 98 & 94 & 119& 161&250\\
		\hline 
	\end{tabular}
}	
}
\end{table}

\section{Conclusions}
\label{sec:conclusions}

In this work, a novel stable third-order accurate numerical method is proposed for micromagnetic simulations. The core of this scheme comprises three key components: a third-order backward differentiation formula (BDF3) approximation for the temporal derivative, a semi-implicit treatment of the cross-product nonlinear term, and a fully explicit extrapolation for the coefficients. Owing to its third-order accuracy, the proposed method demonstrates significantly higher computational efficiency compared to existing lower-order semi-implicit schemes. Comprehensive numerical results from both one-dimensional (1D) and three-dimensional (3D) simulations validate the accuracy and efficiency of the developed method. Furthermore, micromagnetic simulations using the proposed method yield physically consistent microstructures and successfully capture the key dependencies of domain wall velocity, which align with previously reported results.

\section*{Acknowledgments}
This work is supported in part by the Jiangsu Science and Technology Programme-Fundamental Research Plan Fund (BK20250468), Research and
 Development Fund of XJTLU (RDF-24-01-015).

\bibliographystyle{amsplain}
\bibliography{references}

\end{document}